\PassOptionsToPackage{table}{xcolor} 
\documentclass[sigconf]{acmart}

\AtBeginDocument{%
  }

\setcopyright{none}
\acmConference[CCS '25] {Proceedings of the 2025 ACM SIGSAC Conference on Computer and Communications Security}{ October 13--17, 2025}{Taipei}
\acmBooktitle{Proceedings of the 2025 ACM SIGSAC Conference on Computer and Communications Security (CCS '25), October 13--17, 2025, Taipei}
\acmISBN{979-8-4007-1525-9/2025/10}
\acmDOI{10.1145/3719027.3765157}

\usepackage[english]{babel} 
\usepackage{textcomp}
\usepackage[table]{xcolor}
\usepackage{amsfonts}
\usepackage{graphicx}
\usepackage{adjustbox, booktabs}

\usepackage{tikz}

\usepackage[ruled,linesnumbered]{algorithm2e}
\usepackage{balance}

\graphicspath{{./Pics/}}

\DeclareGraphicsExtensions{.jpeg,.png,.pdf}

\usepackage{multirow}
\usepackage[spaces,hyphens]{xurl}
\usepackage{caption, subcaption, adjustbox, booktabs}
\usepackage[ruled,linesnumbered]{algorithm2e}
\usepackage[most]{tcolorbox}
\usepackage[all]{nowidow}
\usepackage[flushleft]{threeparttable}

\usepackage{enumitem}
\setlist[itemize]{leftmargin=6mm}
\usepackage{fontawesome}

\DeclareMathAlphabet{\mathcal}{OMS}{cmsy}{m}{n}

\definecolor{codegreen}{rgb}{0,0.6,0}
\definecolor{codepurple}{rgb}{0.58,0,0.82}
\definecolor{codeblue}{RGB}{0,173,239}
\definecolor{diffblue}{RGB}{15,90,179}

\newcommand{\figref}[1]{Figure~\ref{#1}}
\newcommand{\tabref}[1]{Table~\ref{#1}}
\newcommand{\secref}[1]{Sec.~\ref{#1}}
\renewcommand{\eqref}[1]{Equation~(\ref{#1})}
\newcommand{\appref}[1]{Appendix~\ref{#1}}

\newcommand{\code}[1]{\fontfamily{txtt}\selectfont {#1}\rmfamily}
\newcommand{\dset}[1]{\underline{\textit{#1}}}

\newcommand{\done}[1]{{#1}}

\newcommand{\diffccs}[1]{{#1}}

\newcommand*\circled[1]{\tikz[baseline=(char.base)]{
            \node[shape=circle,fill,inner sep=1pt] (char) {\textcolor{white}{#1}};}}

\newcommand{\colorpar}[1]{\begin{tcolorbox}[colback=gray!10,
                  colframe=black,
                  width=\linewidth,
                  arc=0.7mm, auto outer arc, 
                  boxrule=0.5pt,
                  left=0.7mm,
                  right=0.7mm,
                  top=0.7mm,
                  bottom=0.7mm,
                 ]#1
                 \end{tcolorbox}
                 }

\usepackage{lstautogobble}

\lstdefinestyle{customc}{
  belowcaptionskip=1\baselineskip,
  breaklines=true,
  frame=single,
  numbers=left,
  numbersep=6pt,
  tabsize=2,
  xleftmargin=\parindent,
  language=C,
  showstringspaces=false,
  basicstyle=\fontfamily{txtt}\footnotesize,
  keywordstyle=\bfseries\color{blue!40!black},
  commentstyle=\itshape\color{codegreen},
  identifierstyle=\color{black},
  escapeinside={<@}{@>},
}

\begin{document}

\title{Adversarially Robust Assembly Language Model for Packed Executables Detection}

\author{Shijia Li}
\affiliation{%
  \institution{China Electronics Corporation; Nankai University}
    \city{Shenzhen}
    \state{Guangdong}
  \country{China}
}
\email{lishijia@cec.com.cn}

\author{Jiang Ming}
\affiliation{%
  \institution{Tulane University}
  \state{New Orleans}
  \country{USA}
}
\email{jming@tulane.edu}

\author{Lanqing Liu}
\affiliation{%
  \institution{Nankai University}
    \state{Tianjin}
  \country{China}
}
\email{lqliu@mail.nankai.edu.cn}

\author{Longwei Yang}
\affiliation{%
  \institution{Nankai University}
    \state{Tianjin}
  \country{China}
}
\email{nkuylw@mail.nankai.edu.cn}

\author{Ni Zhang}
\affiliation{%
  \institution{China Electronics Corporation}
      \city{Shenzhen}
    \state{Guangdong}
  \country{China}
}
\email{zhangni@cec.com.cn}

\author{Chunfu Jia}
\authornote{Corresponding author.}
\affiliation{%
  \institution{Nankai University}
  \state{Tianjin}
  \country{China}
}
\email{cfjia@nankai.edu.cn}


\begin{abstract}

Detecting packed executables is a critical component of large-scale malware analysis and antivirus engine workflows, 
as it identifies samples that warrant computationally intensive dynamic unpacking to reveal concealed malicious behavior. 
Traditionally, packer detection techniques have relied on empirical features, such as high entropy or specific binary patterns. 
However, these empirical, feature-based methods are increasingly vulnerable to evasion by adversarial samples or unknown packers (e.g., low-entropy packers). 
Furthermore, the dependence on expert-crafted features poses challenges in sustaining and evolving these methods over time.

In this paper, we examine the limitations of existing packer detection methods and propose \textit{Pack-ALM}, a novel deep-learning-based approach for detecting packed executables. 
Inspired by the linguistic concept of distinguishing between real and pseudo words, we reformulate packer detection as a task of differentiating between legitimate and ``pseudo'' instructions. 
To achieve this, we preprocess native data and packed data into ``pseudo'' instructions and design a pre-trained assembly language model that recognizes features indicative of packed data. 
We evaluate Pack-ALM against leading industrial packer detection tools and state-of-the-art assembly language models. Extensive experiments on over 37,000 samples demonstrate that Pack-ALM effectively identifies packed binaries, 
including samples created with adversarial or previously unseen packing techniques. Moreover, Pack-ALM outperforms traditional entropy-based methods and advanced assembly language models in both detection accuracy and adversarial robustness.

\end{abstract}



\keywords{Malware Analysis; Binary Packing; Assembly Language Model}

\maketitle

\section{Introduction}
\label{sec:introduction}

Malware commonly employs binary packing to evade detection~\cite{Ugarte-Pedrero2015, Mantovani2020, Muralidharan2022}. 
This technique encodes the program's instructions and resources, effectively concealing the original static information from antivirus at a low cost. 
In packed binaries, only the unpacking routine and the compressed (or encrypted) data remain visible to static analysis. 
At runtime, the original instructions are restored and executed by the unpacking routine through complex procedures~\cite{Ugarte-Pedrero2015}.
Furthermore, malware authors often customize packers with various evasion techniques, complicating reverse engineering efforts~\cite{Wressnegger2017}.
As a result, malware researchers primarily depend on static scanning to identify packed binaries that warrant further investigation through dynamic analysis,
which often involves resource-intensive tactics such as hardware-assisted tracing~\cite{Cheng2023} or system-level emulation~\cite{Kawakoya2023}.

\begin{table*}
\centering
\caption{Comparisons of existing packed program detection methods and Pack-ALM. The categories of ``Adversarial Packers?'' refer to three types of modifications employed by adversarial packers: ``Packed Data'' means the modifications involve altering the packing strategy, such as low-entropy packing. ``Metadata'' means the modifications target the alteration of suspicious meta-information such as PE header. ``Instructions'' means the modifications focus on altering entry point and unpacking routine instructions.}
\vspace{-2mm}
\label{tab:detection_classify}

\begin{adjustbox}{width=\linewidth}
\begin{tabular}{lccccccc}
\toprule
\multirow{2}{*}{} & \multirow{2}{*}{\textbf{Detection Target}} & \multirow{2}{*}{\textbf{Feature Granularity}} & \multirow{2}{*}{\textbf{Features \& Threshold}}      & \multirow{2}{*}{\textbf{Novel Packers?}} & \multicolumn{3}{c}{\textbf{Adversarial Packers?}}      
\\ \cmidrule(l){6-8} 
                           &                                          &                                               &             &                                        & Packed Data & Metadata & Instructions \\ \midrule
\textbf{Entropy-based}              & Packed data~\cite{Cozzi2018},~\cite{Lyda07},~\cite{Jeong10},~\cite{entropy17} ,~\cite{erdene13}                           & Whole binary, Section, Window   & {Entropy$\geq$7.0, Entropy$\geq$6.5, Entropy$\geq$7.4}                & \checkmark                                      & \faTimes           & \checkmark                   & \checkmark                 \\
\rowcolor{gray!20}
 & Metadata~\cite{Hai2017},~\cite{2013Malwise},  &  & {Special Strings, PE Header, Import Functions} & & & & \\
 \rowcolor{gray!20}
\multirow{-2}{*}{\textbf{Signature-based}}            & Instructions~\cite{Horsicqc},~\cite{Royal2006},~\cite{Li2023},~\cite{2012SPADE}        & \multirow{-2}{*}{Whole Binary, Specific Address}                & {Byte/Opcode Sequences,  Unpacking Routines}            & \multirow{-2}{*}{\faTimes}                                      & \multirow{-2}{*}{\checkmark}           & \multirow{-2}{*}{\faTimes}                   & \multirow{-2}{*}{\faTimes}                 \\
                   &       Entropy~\cite{Perdisci2008},~\cite{Jacob2013},~\cite{omachi2020packer},~\cite{2016Entropy},~\cite{dam2022packer}  & Whole binary, Section, & {Entropy Patterns, PE Header, N-gram, Strings}                                  &            &                   &                  \\
\multirow{-2}{*}{\textbf{ML-based}}                   & Metadata~\cite{biondi2019effective},~\cite{sun2010pattern},~\cite{santos2011collective}, Instructions~\cite{bergenholtz2020detection},~\cite{hua2020classifying}        & Window, Specific Address &  
{Entry Instruction Bytes, Import Functions, Section } & \multirow{-2}{*}{\faTimes}                                      & \multirow{-2}{*}{\faTimes}           & \multirow{-2}{*}{\faTimes}                   & \multirow{-2}{*}{\faTimes}                 \\
\rowcolor{gray!20}
\textbf{Pack-ALM}                  & Packed data, Instructions                              & Window                  & {Instruction Classification}                                  & \checkmark                                      & \checkmark           & \checkmark                   & \checkmark                 
\\ \bottomrule
\end{tabular}

\end{adjustbox}
\vspace{-4mm}
\end{table*}

Static packer detection has emerged as a promising technique for tackling large-scale packed programs, benefiting both antivirus solutions and malware researchers~\cite{YongWong2021}. 
Existing packer detection methods generally fall into two main categories.
(i) \textbf{Specific Packer Detection}: This approach targets particular packers by identifying unique features, such as metadata in PE headers or byte n-grams~\cite{Aghakhani2020}. 
In pursuit of detection accuracy, security experts often manually extract these features to develop detection signatures or machine-learning (ML) models. 
Identifying distinctive features for specific packed programs typically requires specialized prior knowledge. For example, signature-based tools like YARA~\cite{YaraVictor} and DIE~\cite{Horsicq},
which are employed by VirusTotal~\cite{VirusTotal} and other antivirus, rely on manually crafted rules that require extensive efforts to develop and continuously maintain.
(ii) \textbf{Generic Packer Detection}: This approach aims to detect the presence of packed data, independent of the specific packer used. Existing techniques, including ML models~\cite{Perdisci2008, Jacob2013}, 
are often based on heuristic entropy detection, leveraging the empirical observation that compressed or encrypted data exhibits significantly higher entropy compared to native executables.
Many research studies~\cite{Ugarte-Pedrero2015, Cozzi2018, Lyda07, Jeong10, entropy17} and industrial tools~\cite{Horsicq} use a high entropy threshold (e.g., 7.0 or 6.5) as a key indicator of packed data.

Current packer detection methods struggle to adapt to the evolving landscape of adversarial and customized packers.
Specific packer detection methods are limited to identifying known packers, rendering them ineffective against novel or modified packer variants. 
Their limitations are evident in the slow responsiveness to the packer's specialized modifications.
For instance, TrickGate, a widely used Packer-as-a-Service, has continuously evolved, successfully evading detection for over six years \cite{Olshtein}.
Furthermore, recent advancements in general-purpose unpacking methods~\cite{Cheng2018, Cheng2021, Cheng2023} have further undermined the effectiveness of specific packer detection techniques. 
These approaches aim to recover unpacked programs regardless of the specific packer employed. 
As such, generic packer detection emerges as a promising direction for developing more robust malware detection and analysis strategies.
However, existing entropy-based generic detection frameworks are inherently vulnerable to adversarial techniques such as low-entropy packed samples~\cite{Mantovani2020}.
In summary, current packer detection methods are confronted with the following two problems.

\vspace*{2pt}
\noindent \textbf{P1: Existing packer detection techniques rely heavily on the subjective selection of features and thresholds, which are largely dependent on the analyst's experience.}
For a long time, packer detection methods have been grounded in heuristics-based approaches. For example, entropy-based detection operates on the empirical assumption that entropy levels above 7.0 indicate the presence of packed programs~\cite{Ugarte-Pedrero2015}. Similarly, both rule-based and machine-learning detection models frequently utilize empirical features derived from program metadata, such as PE headers and section information~\cite{Li2022, Mantovani2020, DHondt2024}.
Unfortunately, the reliance on expert judgment to select detection features introduces significant limitations, given the inherent subjectivity and incompleteness of human knowledge. 
A recent study~\cite{DHondt2024} shows that attackers can easily camouflage these features to bypass packer detection systems that rely on entropy or metadata signatures. 
Additionally, Mantovani et al.~\cite{Mantovani2020} demonstrates that more than 30\% of malware samples employ multiple encoding tricks to artificially reduce their entropy levels below the common threshold of 7.0.

\vspace*{2pt}
\noindent \textbf{P2: The granularity of empirical detection techniques directly impacts the accuracy of packer detection and their resilience to adversarial techniques.}
Packed programs often utilize diverse strategies to conceal packed data within existing sections (e.g., ``\code{.text}'' or ``\code{.rsrc}'') or newly created sections. 
Consequently, entropy-based approaches may fail to detect packed data signals due to their dependence on manually defined levels of granularity (e.g., whole program, individual sections, or sliding windows)~\cite{Ugarte-Pedrero2015, Mantovani2020}. 
Our findings indicate that the choice of granularity significantly affects detection accuracy. 
For example, entropy detection at the section level can increase false positives by up to 818.2\% (as detailed in \secref{sec:sec3entropy_eva}).
Similarly, the choice of an appropriate detection range also poses challenges to the packer detection rule developers.
Li et al.~\cite{Li2023} report that over 93\% of human-crafted packer detection rules only scan at specific addresses, rendering them susceptible to circumvention by malware authors.

\tabref{tab:detection_classify} summarizes the existing packer detection techniques in terms of detection objects, feature granularity, and resistance to packer variants (i.e., novel and adversarial packers).
Notably, entropy-based methods demonstrate considerable generality against novel and adversarial packers when directly targeting packed data, even when these packers employ modified metadata and instructions. 
However, the susceptibility of entropy metrics to adversarial low-entropy encoding techniques remains a significant limitation. 
Therefore, advancing robust detection methods for packed data continues to be a valuable and promising area.

Leveraging language models (e.g., BERT~\cite{Devlin2019}) from the field of natural language processing (NLP), recent studies~\cite{Pei2021, Li2021, Jin2022, Wang2022, Luo2023} 
have demonstrated notable success in various binary analysis tasks, such as disassembly. However, no prior research has investigated the effectiveness of these models in handling real-world code obfuscation techniques, 
including binary packing~\cite{UPXTeam} and code virtualization~\cite{Softwarea, Technologiesa}. For example, binary packing often employs complex instruction combinations and interleaves instructions with packed or junk data, 
which can potentially disrupt assembly language models trained on standard instruction sequences.

In this paper, we introduce \textit{Pack-ALM}, a novel \emph{Pack}ing-aware \emph{A}ssembly \emph{L}anguage \emph{M}odel designed to address the aforementioned limitations and advance packer detection.
Inspired by the principles of grammatical error correction and real-word/pseudoword distinction in linguistics~\cite{Thomas1997}, Pack-ALM identifies the packed data by distinguishing ``real'' and ``pseudo'' instructions, 
which are generated through linear-sweep disassembly.
This transformation reframes the packer detection challenge as a problem of identifying assembly instructions with syntax and format errors, thereby aligning the task with the capabilities of NLP models. 
Pack-ALM classifies programs (or areas) based on the proportion of identified packed data, as such data serve as a distinguishing feature that differentiates packed programs from non-packed ones.

Specifically, we built Pack-ALM on top of the RoBERTa model~\cite{Liu2019}, incorporating a novel approach to instruction preprocessing and a refined masked language modeling (MLM) task.
This design enables Pack-ALM to effectively distinguish between real instructions and packed data.
First, we perform a linear disassembly of non-packed programs and preprocess the resulting data into normalized instruction representations. 
We then employ the refined MLM task to pre-train Pack-ALM, enabling it to capture the structural components of instructions. 
Subsequently, Pack-ALM is fine-tuned to classify pseudo instructions derived from packed data versus real instructions.
For the downstream application of identifying packed programs (or areas), we feed the normalized instructions as the input of Pack-ALM and summarize the detection results.

To evaluate the efficacy of Pack-ALM, we conducted a series of experiments using over 37K samples, comprising packed samples generated by both widely used real-world packers and custom adversarial packers (including low-entropy variants), as well as non-packed samples.
Our evaluation results demonstrate that Pack-ALM outperforms existing state-of-the-art assembly language models~\cite{Li2021,Pei2021,Wang2022} as well as signature-based and
entropy-based tools~\cite{Horsicq,Li2023}. Our approach achieves significantly higher precision, recall, and F1 scores than other models in identifying packed data, including samples from previously unseen packers and low-entropy packed programs.
Furthermore, Pack-ALM can be extended to robustly classify real-world non-packed and packed programs. Extensive evaluations in real-world scenarios indicate that Pack-ALM reliably identifies packed data at a fine granularity. 
It consistently outperforms peer tools in resisting adversarial techniques, such as sophisticated obfuscation and adversarial packing.

\vspace*{2pt}
\noindent \textbf{Contributions }
Our key contribution is to investigate the inherent weakness of the long-standing entropy metrics and propose a novel detection method to
complement existing packer detection techniques for the security community.
\diffccs{The proposed Pack-ALM model enables precise identification of packed data, along with its location within packed programs. 
This capability empowers analysts to effectively track and monitor packed regions during both static and dynamic analyses. }
In summary, our paper makes the following technical contributions:

\begin{itemize}
    \item We evaluate and highlight the intrinsic limitations of long-standing entropy metrics. 
    \item We introduce a novel approach to packer detection by reframing the challenge as a problem of distinguishing real instructions from pseudo ones. 
    \item We propose a single unsupervised task to pre-train a packing-aware assembly language model, Pack-ALM, for identifying packed regions and packed programs. It exhibits resilience against binary packing and complex code obfuscation.
    \item We conduct large-scale evaluations and reveal the superiority of Pack-ALM to other assembly language models in packer detection, particularly in recognizing both novel and adversarially packed programs.
\end{itemize}

\noindent \textbf{Open Source }
We release Pack-ALM's source code and pre-trained model to facilitate reproduction and reuse,
as all found at \href{https://zenodo.org/records/14091136}{\underline{Zenodo}}. 

\section{Background, Related Work, and Motivation}
\label{sec:background}
This section lays the groundwork for our packing-aware assembly language model. We begin by reviewing existing research on binary packing and corresponding detection techniques. 
Then, we delve into a multi-dimensional analysis of entropy metrics and highlight the inherent limitations of entropy-based detection for packed program detection. 
This analysis motivates our exploration of adversarially robust assembly language models. 

\subsection{Binary Packing}
\label{sec:sec2binary-packing}

This subsection introduces the workflow of binary packing, focusing on the static composition of packed programs. 
It investigates the content differences between packed and non-packed programs, and analyzes these discrepancies in the context of existing packer detection techniques.

\begin{figure}
    \setlength\abovecaptionskip{-0.001\baselineskip}
    \centerline{\includegraphics[width=\linewidth]{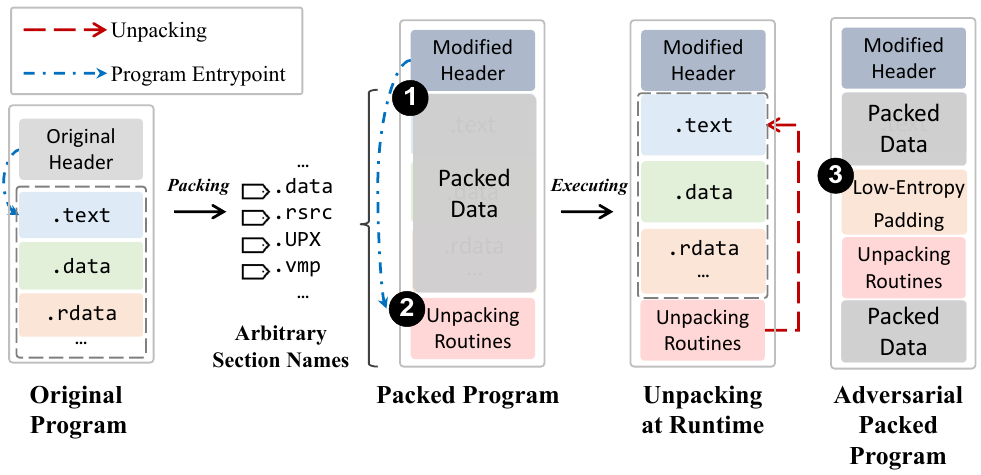}}
    \caption{\label{fig:05-static-packer} Overview of binary section changes during the packing/unpacking process.}
    \vspace{-5mm}
\end{figure}

\figref{fig:05-static-packer} illustrates the content modifications introduced during the packing process of a typical program. The packed program comprises two key components: packed data (\circled{1} in \figref{fig:05-static-packer}) and the unpacking routine (\circled{2}). 
Packed data consists of unordered bytes generated by compressing (or encrypting) the original program's code and essential resources. 
These packed data are typically stored in large, consecutive segments within the packed program. 
The unpacking routine is the assembly instructions, which are used to sequentially decompress (or decrypt) the packed data at runtime and enable the released original instructions to be executed normally~\cite{Pericin2011,Corporation,Barabosch}.
To hinder static analysis, the unpacking routines are frequently obfuscated and intermixed with the packed data.

Compared with standard programs, packed programs exhibit significant structural differences in terms of program content organization. Standard programs typically store executable instructions within the ``\code{.text}'' section, whereas packed programs may disperse unpacking routines and packed data across various sections, even in the ``\code{.rsrc}'' resource section.
To evade detection, packers often arbitrarily modify header information and non-essential program segments without hindering the packed program's execution or functionality. 
To counteract potential evasion techniques, packed program detection often necessitates a comprehensive scan of the entire program.

Furthermore, the assembly instructions of packed programs are often heavily obfuscated and lack clear function and instruction boundaries~\cite{Li2023}, posing significant challenges for disassembly tools in accurately distinguishing real instructions from packed data. For instance, PECompact-packed programs employ obfuscated instructions to intentionally trigger faults, with control subsequently redirected to the actual unpacking routines via registered exception handlers. When analyzing such packed programs, even state-of-the-art commercial disassembly tools like IDA Pro fail to correctly identify instruction boundaries, resulting in the misidentification of 109 instructions (as detailed in \appref{app:case_study1}).

Additionally, disassembly methods based on superset disassembly~\cite{Bauman2018}, such as those proposed in~\cite{Yu2022, Miller2019}, also encounter substantial difficulties due to the high volume of false-positive instructions generated from packed code. For example, when superset disassembly is applied to ZProtect-packed programs, the average false positive rate reaches as high as \done{99.5\%} (discussed in \appref{app:redundant-disassembly}).

\subsection{Classification of Packer Detection}
\label{sec:packer-detection}
We surveyed popular packer detection tools and the papers published in \done{18} major cyber security venues over the past \done{20} years. 
As summarized in \tabref{tab:detection_classify}, the existing packer detection techniques primarily rely on the subjective choice of features and granularity.
Signature-based and ML-based packer detection methods typically scan the entropy, metadata (e.g., PE header information) and disassembled instructions at specific addresses (e.g., the entry point) or the entire binary.
\diffccs{
However, these feature-based detection mechanisms render them inherently vulnerable to adversarial techniques, such as low-entropy packed samples.
State-of-the-art tools like PackGenome~\cite{Li2023} (signature-based) and PackHero~\cite{packhero_2025} (ML-based) fail to address this comprehensively.
Although PackGenome was evaluated on low-entropy datasets, its performance was limited: it detected few off-the-shelf packed samples and failed to identify customized packers.
Similarly, PackHero demonstrates efficacy only for small, low-entropy packed samples exhibiting specific features.
}

\diffccs{Furthermore}, their reliance on manually defined features makes them struggle to adapt to the detection of novel and adversarial packers.
For example, malware authors often customize the open-sourced packer UPX with diverse entry point instructions and unpacking routines.
Consequently, the detection tools need to continuously update their features and signatures to identify customized variants~\cite{DHondt2024,Li2023}.

The entropy-based methods treat the program (or data) with high entropy as the indication of packing.
Since the packed data is the essential composition of the packed program, entropy-based detection is promising to identify the diverse packed program, such as novel and adversarial packed programs.
However, this method is greatly influenced by the empirically defined granularity and is vulnerable to the low-entropy evasion techniques~\cite{Mantovani2020}.

\subsection{Entropy-Based Packer Detection}

Entropy analysis has long been a foundational technique for generic packer detection. The core principle is that non-packed programs, which contain both instructions and data,
exhibit an inherent order and therefore display relatively low entropy. In contrast, packed data---often compressed or encrypted---typically shows higher entropy~\cite{Lyda07}. 
Compression seeks to represent information more compactly, while encryption maximizes data disorder. Based on this empirical observation, entropy-based detection methods identify 
high entropy as an indicator of packed programs. However, existing methods vary in their choice of entropy threshold and detection granularity.

\vspace*{2pt}
\noindent\textbf{Entropy Threshold Selection } Most studies, including state-of-the-art academic works~\cite{Ugarte-Pedrero2015} and industry tools (e.g., Manalyze~\cite{JusticeRage}), have established a common heuristic: a program's entropy exceeding 7.0 indicates potential packing. However, this threshold varies across detection tools to balance the trade-off between false positives and false negatives. 
For example, DIE~\cite{Horsicq} adopted a more aggressive threshold to enhance adversarial sample identification, lowering the standard threshold from 7.0 to 6.5 as of version 2.05~\cite{Horsicqc}. 
In contrast, Malscan~\cite {Malscan} and pefile~\cite{Erocarrera} apply a higher threshold of 7.4 to reduce false positives.

\vspace*{2pt}
\noindent\textbf{Three Levels of Detection Granularity }
The selection of the input region for the detection model can be categorized into three levels of granularity: whole program, section, and sliding window. 
While whole-program granularity was commonly used in earlier research, its coarseness makes it vulnerable to targeted evasion techniques.
To address this, Mantovani et al.~\cite{Mantovani2020} and popular tools like Manalyze~\cite{JusticeRage} perform section-based scanning. 
For example, pefile \cite{Erocarrera} classifies a program as packed if sections with high entropy constitute more than 20\% of the total number of sections. 
For even finer detection, Kawakoya et al.~\cite{Kawakoya2023} propose dividing program code into overlapping windows and selectively classifying them based on entropy variations.

However, current practices in choosing entropy thresholds and detection granularity largely depend on researchers' subjective experience and lack systematic large-scale validation. 
This reliance on heuristic choices can lead to elevated false positive or false negative rates in real-world scenarios.

\subsection{Limitations of the Entropy-based Generic Packer Detection}
\label{sec:sec2challenges_generic}

The rapid evolution of customized packed programs and the growing demand for malware analysis have exposed the limitations of traditional, specific packer detection methods, which struggle 
to keep up with the emergence of novel and adversarial packers. Moreover, recent advancements in generic binary unpacking~\cite{Cheng2018,Cheng2021,Cheng2023,Kawakoya2023} 
have rendered specific packer detection tools increasingly obsolete.

In contrast, generic packer detection provides a more scalable solution, particularly suited for large-scale binary unpacking scenarios. However, existing generic detection approaches rely heavily on entropy-based metrics, which require expert-defined thresholds and granularity settings. This reliance introduces two primary limitations:

\vspace*{2pt}
\noindent\textbf{L1: Entropy-based metrics fail to capture the semantic content of data.}
While sensitive to data distribution, information entropy neglects the arrangement of data elements (detailed in \secref{sec:sec3entropy_eva}). 
Low-entropy packing techniques can bypass entropy-based detection by preserving the original data distribution.
For example, mono-alphabetic substitution can produce packed data with entropy identical to that of the original instruction sequence~\cite{Mantovani2020}, 
as this substitution merely rearranges bytes without altering the overall data distribution.
Furthermore, entropy represents only statistical disorder of data and lacks the capacity to capture semantic characteristics, such as instruction patterns and data structures. 
This limited detection scope often necessitates the combination of entropy with additional empirical indicators (e.g., entry-point instructions) in YARA rules and machine learning models for generic packer detection.

\vspace*{2pt}
\noindent\textbf{L2: Entropy-based generic packer detection heavily relies on expert experience.} 
Existing entropy-based detection methods rely on two key parameters: the entropy threshold and the scan granularity. 
Both parameters are typically determined through expert judgment, without systematic evaluation. 
Current studies employ varying subjective entropy thresholds (e.g., 7.0 or 7.4) to differentiate between packed and unpacked regions, introducing inconsistency in defining ``high entropy,'' 
which can significantly impact detection performance. Traditional thresholds (e.g., entropy $H(X) \geq 7.0$) allow low-entropy packing techniques to evade detection, whereas overly low thresholds (e.g., $H(X) \geq 6.5$) result in numerous false positives.
Similarly, the selection of the scan granularity remains an experience-based practice. Conventional methods often target the entire program or specific sections, 
but they have proven ineffective against byte-padding techniques employed by low-entropy packers~\cite{Mantovani2020} (\circled{3} in \figref{fig:05-static-packer}). 
Conversely, excessively fine-grained detection can lead to high false positives (detailed in Appendix~\ref{sec:sec3entropy_eva}).

To address these limitations, a promising direction is the \diffccs{use of scanning techniques} to detect packed data within packed programs. 
However, to improve robustness against novel and adversarial packers, such approaches should avoid heuristic algorithms that depend heavily on expert knowledge.

\subsection{Assembly Language Models}
\label{sec:sec2assembly-model}

In recent years, the transformer model~\cite{Vaswani2017} has gained widespread adoption across various fields. 
For example, ViT~\cite{Dosovitskiy2021} enables ``natural language-like'' image processing by transforming images into formats that resemble word sequences. 
A critical component of this process involves appropriately tailoring input preprocessing and modifying the transformer model's structure to incorporate domain-specific knowledge.
{Numerous studies have explored the adaptation of the transformer model to binary analysis workflows.}

{
\noindent \textbf{XDA}~\cite{Pei2021} employs a self-supervised pretraining task to capture contextual dependencies among byte sequences in binaries, generating byte embeddings for downstream tasks like function boundary detection and instruction recovery.
}

{
\vspace*{2pt}
\noindent \textbf{PalmTree}~\cite{Li2021} models assembly instructions analogously to natural language sentences. This approach uses three pre-training tasks to learn the contextual relationships by capturing both control flow and data flow dependencies between instructions. The pre-trained model can be applied to downstream tasks such as binary code similarity detection. 
}

{
\vspace*{2pt}
\noindent \textbf{jTrans}~\cite{Wang2022} captures structural dependencies between direct jump operations by employing parameter-shared embeddings. The model is trained using a self-supervised objective that combines masked language modeling with jump target prediction. By incorporating control flow semantics into both the embedding representations and the pre-training objective, jTrans enhances the effectiveness of binary code similarity detection.
}

{
\vspace*{2pt}
\noindent \textbf{CLAP}~\cite{wang2024clap} enhances the complex semantic representation and transferability of learned representations by aligning assembly code with corresponding natural language explanations. However, this approach misaligned with packing-specific challenges, where code obfuscation disrupts structural and semantic relationships between instructions and their textual explanations.
}

Unfortunatey, code obfuscation poses significant challenges for current assembly language models. Existing models~\cite{Pei2021,Li2021,Wang2022} are typically trained on open-source program datasets (e.g., Byteweight~\cite{Bao2014} and SPEC benchmark~\cite{SPEC}), 
where clear boundaries exist between instruction and data sections. In contrast, commercial programs and malware frequently employ sophisticated obfuscation techniques, such as binary packing, that extensively interweave instructions with packed data.
For example, packed programs often use aggressive control flow obfuscation methods, eliminating function boundaries and embedding instructions and data within non-standard sections beyond ``\code{.text}''. This disrupted arrangement deviates from standard function and control flow patterns, posing significant challenges for both disassembly tools and models. In particular, approaches~\cite{Yu2022} based on superset disassembly~\cite{Bauman2018}, which sequentially disassemble all bytes of a program, are prone to generating excessive redundant instructions from packed data.

\subsection{Scope of Pack-ALM}
\label{sec:sec2motivation}
\diffccs{
Detecting packed programs, especially adversarial or novel variants, remains a persistent challenge. To address this, we introduce Pack-ALM, a novel methodology designed to enhance the detection of packed programs, with particular efficacy against adversarial packers.
Pack-ALM requires training on only a limited set of packers. This enables robust identification of a broad spectrum of similar low-entropy and novel samples. Pack-ALM offers two key advantages:
}

\begin{enumerate}
    \item \diffccs{
Enhanced Reverse Engineering: Pack-ALM can significantly improve static analysis by identifying packed regions and highly obfuscated instructions, surpassing tools like IDA Pro~\cite{idapro}.}
    \item \diffccs{Streamlined Malware Analysis: Pack-ALM directly assists state-of-the-art malware analysis tools, such as XunPack~\cite{Kawakoya2023}, by providing localization of packed data—a critical step for reconstructing the original malicious code.
}
\end{enumerate}

 \begin{figure*}[htp]
    \centerline{\includegraphics[width=0.95\textwidth]{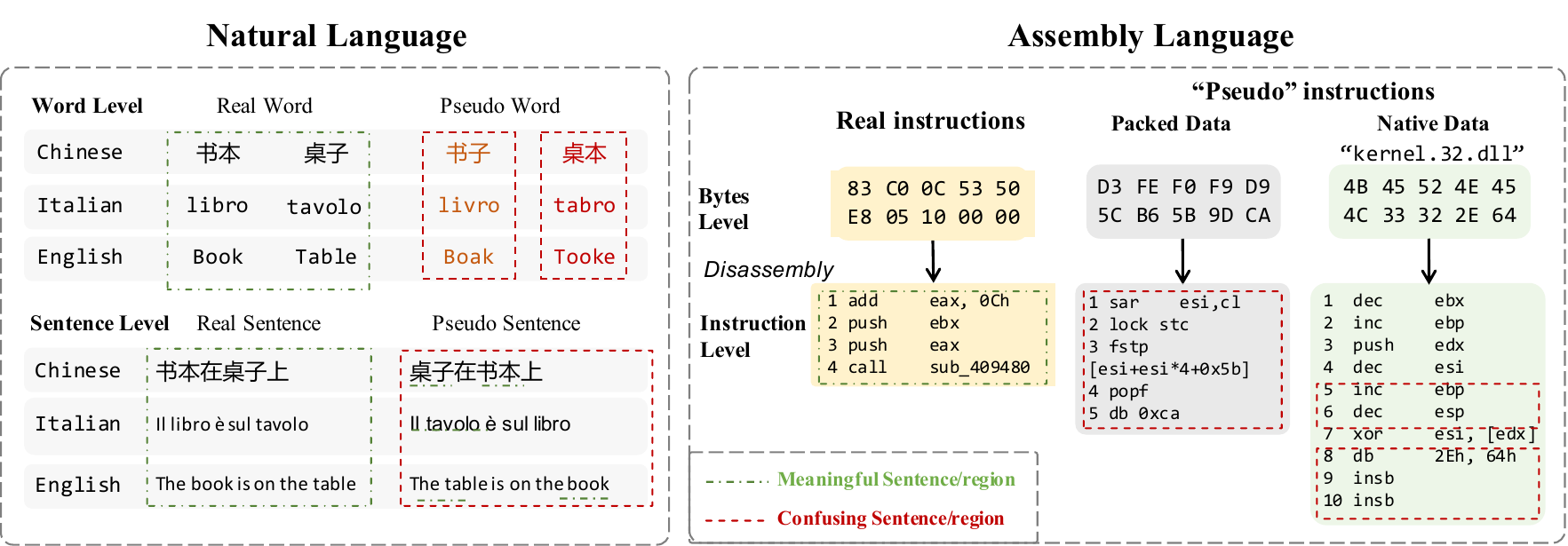}}
    \vspace{-2mm}
    \caption{\label{fig:04-nature-assembly} An analogy between real-word/pseudoword distinctions in natural language and real/pseudo instruction distinctions in packed executables.}
    \vspace{-5mm}
\end{figure*} 
\section{Our Approach}
\label{sec:sec4approach}


\subsection{The Representation of Binary Program}
\label{sec:sec4representation}

Adapting natural language models (e.g., Transformer) to the binary analysis domain requires first converting binary programs into a suitable representation 
and tailoring the pre-training task for understanding the representation. 
In the specific context of detecting packed programs or packed data, the model must be capable of distinguishing among real code, native data, and packed data.

Drawing inspiration from linguistic studies on word distinction, we propose a novel approach for representing and distinguishing program content to mitigate the aforementioned challenges.
In linguistics, a ``real word'' possesses a well-defined meaning, while a ``pseudoword'' is a meaningless string that is pronounceable and looks like real words.
Akin to natural language, instructions disassembled from the executable code can be considered as ``real instructions'', whereas instructions converted from native data or packed data can be seen as ``pseudo instructions.''
As illustrated in \figref{fig:04-nature-assembly}, we covert the packed data detection problem into two steps: 1) differentiating ``pseudo instructions'' from ``real instructions,'' and 2) 
further classifying among types of ``pseudo instructions.''

The distinction between the real words and the pseudowords serves as a metric for evaluating the ability to comprehend and discern word semantics. 
For example, humans demonstrate faster reaction times to real words like ``book'' compared to pseudowords like \textcolor{orange}{``boak''}.
Moreover, humans generally respond more rapidly and accurately to pseudowords that closely resemble real instructions (e.g., \textcolor{orange}{``boak''}) compared to those that bear less similarity (e.g., \textcolor{red}{``tooke''})~\cite{Thomas1997}.

Similarly, reverse engineers exhibit quicker responses to real instructions than pseudo instructions (``\textcolor{red}{Confusing Sentence/Region}'' in \figref{fig:04-nature-assembly}).
While both native and packed data within packed programs fall under the category of pseudo instructions, the instructions converted from native data exhibit a greater resemblance to real instructions.
This can be attributed to the more organized nature of native data compared to packed data, which is reflected in the disassembled instructions.
For example, instructions extracted from the human-readable string ``\code{kernel32.dll}'' in \figref{fig:04-nature-assembly} retain specific combination patterns, demonstrating a higher degree of organization than those extracted from packed data.

Furthermore, existing linguistic research~\cite{PICKERING1999136} suggests that well-designed contexts and syntactic patterns can trigger semantic and syntactic priming effects, thereby enhancing subsequent sentence processing and comprehension. 
For example, the sentence ``doctor eats bread'' is generally understood more rapidly and accurately than ``bread eats doctor'' due to its alignment with common knowledge and syntactic conventions.
Similarly, the reverse engineers can easily identify the function call within \figref{fig:04-nature-assembly}, but encounter confusion when analyzing pseudo-instructions.

{This structural resemblance motivates the application of linguistic research principles to improve model design and assembly language representation. This theoretical foundation also supports our methodology for binary instruction classification tasks. 
Our approach aims to facilitate the model's ability to naturally learn distinctions among instructions, native data, and packed data.}

\subsection{Pack-ALM Overview}

\begin{figure*}
    \setlength\abovecaptionskip{-0.001\baselineskip}
    \centerline{\includegraphics[width=\textwidth]{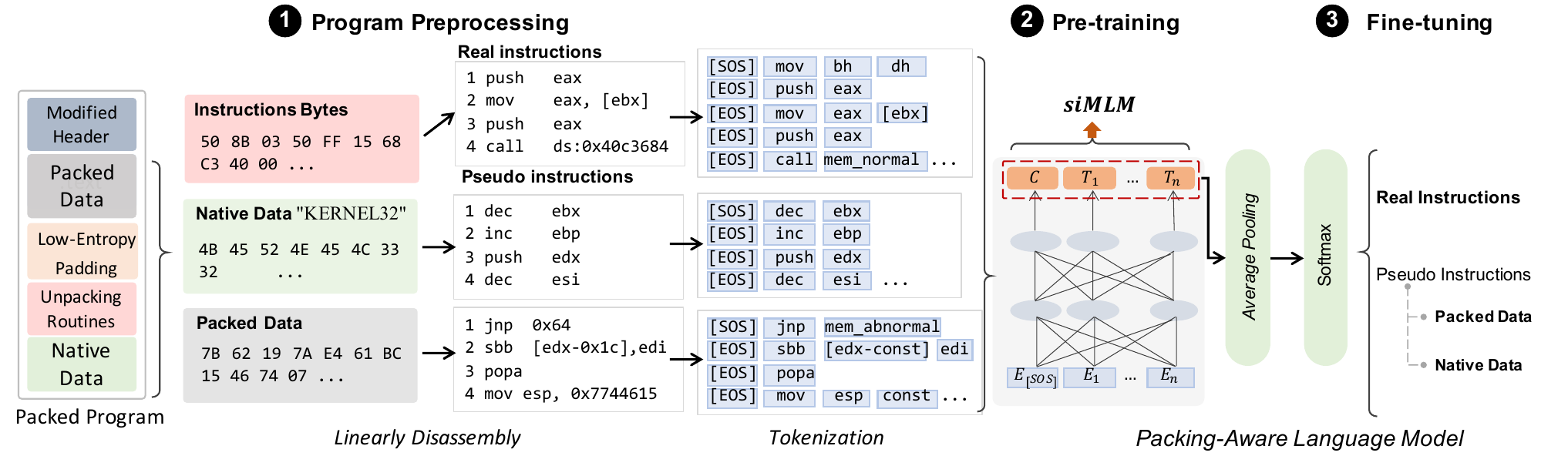}}
    \caption{\label{fig:05-pack-alm} Overview of Pack-ALM model's workflow.}
    \vspace{-3mm}
\end{figure*}

To address  the limitations of entropy-based generic binary packing detection methods and reduce reliance on expert knowledge, we introduce Pack-ALM, a language model designed to identify packed data regions within binary programs. 
Pack-ALM differentiates packed data from both instructions and native data, thereby enabling the recognition of packed regions and programs. 
This approach draws inspiration from linguistic research that distinguishes between real words and pseudowords. We propose a novel instruction preprocessing method and refine the pre-training tasks to enhance the model's resilience to adversarial techniques, such as obfuscation and low entropy packing.
The architecture of Pack-ALM is depicted in \figref{fig:05-pack-alm}. 
Its core functionality involves three primary processes:


\vspace*{2pt}
\noindent\textbf{\circled{1} Program Preprocessing } 
We begin by applying linear disassembly to the raw data of the input (packed) programs, disassembling both code and data regions into instruction sequences. 
The objective of using linear disassembly is not to achieve ``perfect disassembly''---a task inherently infeasible for obfuscated or packed code---but rather to expose structural 
irregularities by interpreting packed data as instructions.
To improve the model's ability to distinguish between real and pseudo-instructions, we implement a lightweight normalization strategy during instruction preprocessing. 
After tokenizing the instructions, we divide the resulting token sequences into regions, which are then used as input to the model.

\noindent\textbf{\circled{2} Pre-trained Model Construction } 
Inspired by linguistic studies, we propose a complementary pre-train task: {Structured Instruction MLM (siMLM) task to learn compositional relationships (e.g., test-je sequences)}. 
This siMLM is designed to learn the compositional relationships among tokens and combinations of instructions.
To enhance the model's ability to capture syntactic relationships between assembly instructions, we use a dataset of non-packed programs for pre-training.

\vspace*{2pt}
\noindent\textbf{\circled{3} Model Fine-Tuning }
To fine-tune the pre-trained model, we first employ average pooling to downsample its output. 
Then, an additional dense layer followed by a softmax classifier is incorporated.
The softmax classifier categorizes each input region into one of the following labels: instruction, native data, or packed data.
During the fine-tuning process, we use a dataset comprising packed programs, containing obfuscated instructions and packed data. 
Packed programs represent more realistic scenarios encountered in real-world packed program detection, as they exhibit sophisticated code obfuscation. 
This fine-tuning enables the model to acquire the necessary knowledge for downstream tasks.

\subsection{Program Preprocessing}
\label{sec:sec4preprocess}

As discussed in \secref{sec:sec2binary-packing}, obfuscation techniques often disrupt function boundaries and intermix instructions with data in packed programs.

To overcome the limitations of existing methods and develop robust models without relying on prior knowledge, 
our preprocessing strategy preserves the semantics of instructions while transforming them into a representation suitable for downstream tasks. 
We first employ linear disassembly to convert both the program's code and data sections into instructions.
Then, we directly use instruction prefixes, mnemonics, and register-type operands as tokens. Given their inherent meanings within the Intel instruction set, no modifications are required.
Note that obfuscation techniques can manipulate the EFLAGS register to complicate analysis~\cite{Li2022} (e.g., VMProtect uses redundant \code{bts} and \code{cmc}).
Therefore, we omit instructions' implicit EFLAGS register operand.

Given that the model's primary objective is to identify packed regions, 
analyzing the immediate operands of function parameters or variable sizes is unnecessary.
To enhance the model's robustness in distinguishing between real and pseudo instructions, we implement the following modifications of the immediate operand during the normalization stage:

\begin{itemize}
\item
    For arithmetic operation instructions such as \code{add eax, 0x1}, we directly mark the immediate operand as ``\code{[const]}.'' 
\item
    For memory operation instructions like \code{mov eax, [eax+IMM]}, we assign labels to immediate operands based on the validity of the memory region. Specifically, immediate operands within the program's normal memory range are labeled as ``\code{[const\_normal]},'' while those outside the program's memory range are labeled as ``\code{[const\_abnormal]}.'' For example, an operand like ``\code{[eax-0x6281719]}'' is considered an abnormal constant as it exceeds the program's memory boundaries.
\item
    For addressing operation instructions such as \code{jmp IMM}, we assign labels to immediate operands according to the validity of the target address. A target address within the program's valid address range is 
    labeled as ``\code{[mem\_normal]},'' while those outside this range are labeled as ``\code{[mem\_abnormal]}.''
\item
    For unparsable data that cannot be disassembled, we label the data commonly used for padding (e.g., 0x00) in compiled programs as ``\code{[pad\_normal]},'' and other data as ``\code{[pad\_abnormal]}.''
\end{itemize}

\diffccs{This approach reduces vocabulary complexity by preventing excessive token set expansion, and enhances generalizability by focusing on universal packed data characteristics rather than packer-specific numeric constants. By excluding immediate values, the model learns transferable patterns distinguishing from unpacked instructions across diverse packing techniques. }

Then, we employ special tokens \code{[SOS]} and \code{[EOS]} to denote the beginning of the whole sequence and separator of each instruction, respectively.
Finally, these preprocessed instructions are assembled into windows with a maximum length of 512 tokens as the input of our model.

During the preprocessing stage, immediate operands are labeled based on the program's memory scope, removing potential bias introduced by human-defined heuristics. 
This strategy also mitigates the out-of-vocabulary (OOV) problem while preserving the semantics of assembly instructions.

\subsection{Pack-ALM Model}
\label{sec:sec4packalm}

Pack-ALM builds on the RoBERTa model \cite{Liu2019}, which enhances the BERT model by optimizing pre-training tasks and hyperparameters. Notably, RoBERTa has shown that the Next Sentence Prediction task offers minimal benefit to model performance. Consequently, Pack-ALM exclusively utilizes a Masked Language Modeling (MLM) pre-training task to capture the compositional and contextual relationships among assembly instructions.

\subsubsection{Component-Aware Masked Language Model Task}

\label{sec:sec4component-aware}
We introduce a {structured instruction} MLM training task, denoted as siMLM, to learn compositional relationships within assembly instructions.
Our siMLM task enforces a mutually exclusive random masking and replacing strategy at the instruction component level. 
The mutually exclusive principle is based on the instruction construction and opcode-operand combination.
This approach aligns with the intuition of reverse engineers, who learn the functionality of assembly instructions by considering the combinations of opcodes and operands.

Given an input instruction sequence, we randomly select 20\% of the tokens for replacement. Among the selected tokens, 40\% of mutually exclusive tokens are assigned a special \code{[MASK]} token, 50\% are replaced with randomly chosen mutually exclusive tokens from the dictionary, and the remaining 10\% are kept unaltered.
Additionally, the \code{[EOS]} token is not masked, as our model does not need to identify instruction boundaries.
To help the model learn instruction components, our mutually exclusive principle prevents simultaneous masking of both the opcode and a portion of the operand.
For example, given the instruction ``\code{mov eax, ebx}'', non-mutually exclusive masking it as ``\code{[MASK] [MASK] ebx}'' would hinder the model's ability to predict both the opcode and operand from the masked operand.
The model then predicts the masked lexical elements, the replaced tokens, and the unaltered tokens, and outputs the predicted probability for the masked token.

\begin{figure}
    \centerline{\includegraphics[width=\linewidth]{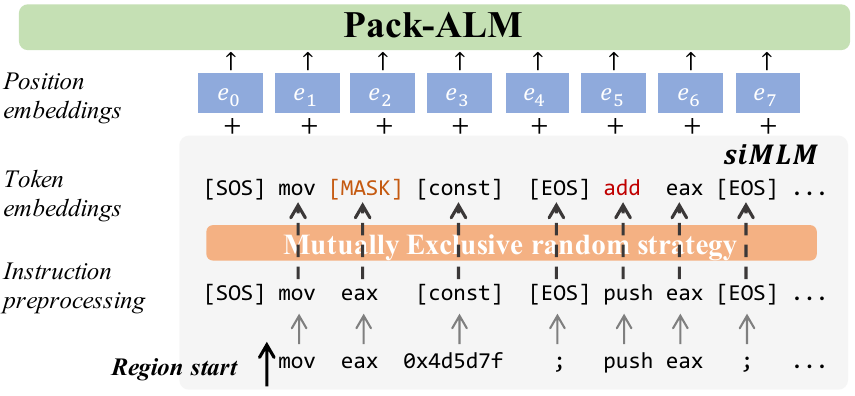}}
    \caption{\label{fig:04-model-masking} Component-Aware Masked Language Model.}
    \vspace{-5mm}
\end{figure}

The training process uses a cross-entropy loss function. We consider the input to be a region comprising tokens of assembly instructions, denoted as $R=[t_1,...,t_n]$, where $t_i$ represents the $i-th$ token within $R$ and $n$ signifies the total number of tokens. The loss function is:
\begin{equation}
    \mathcal{L}_{siMLM} = - \sum_{t_i\in m(R)} log p(\hat{t_i})
\end{equation}
where $m(R)$ denotes the set of {selected} tokens within sequence $R$. 
\figref{fig:04-model-masking} illustrates an example of masked tokens, 
where the first instruction's \code{eax} is masked and the second instruction's \code{push} is replaced with \code{add}. Our mutually exclusive masking strategy prevents masking both \code{mov} and \code{eax} simultaneously.

\subsubsection{Model Pre-training}
\label{sec:sec4model-pre-training}
For each non-packed program within a given dataset, we linearly disassemble its raw data (not including the program's header) and preprocess the instructions according to the strategy outlined in \secref{sec:sec4preprocess}.
Then, we feed the preprocessed token sequences into the model and employ pre-training tasks siMLM.
To enhance the model's generalization capabilities, we implement a dynamic masking strategy during the pre-training process. This strategy masks different components of each instruction and different instructions in multiple training epochs.

\subsubsection{Model Fine-tuning}
\label{sec:sec4model-fine-tuning}
Fine-tuning refines the pre-trained model's adaptability to the specific task of packed program detection. 
We fine-tune the model using a dataset of packed programs, each preprocessed according to the strategy outlined in \secref{sec:sec4preprocess}. 
To represent the semantic information of a region, we apply average pooling to the embedding vectors, yielding a general semantic representation of the instruction {region}.
After generating the {semantic representation} with the pre-trained model, we add a dense layer followed by a softmax classifier to categorize each region $R$ into one of three classes: instructions, native data, or packed data. 
As this classification task only requires distinguishing region $R$ types, precise disassembly or fine-grained data classification (e.g., identifying jump tables) is unnecessary.

\section{Evaluation}
\label{sec:sec5evaluation}
In this section, we evaluate Pack-ALM by answering the following four research questions (RQs):
\begin{itemize}
    \item 
\textbf{RQ1}: How does Pack-ALM compare to the state-of-the-art models \diffccs{in packed data detection}? 
    \item 
\textbf{RQ2}: Can Pack-ALM generalize to detect model-unseen or adversarial packers? 
    \item 
\textbf{RQ3}: How do Pack-ALM's components contribute to the effectiveness? 
    \item 
\textbf{RQ4}: How effective is Pack-ALM in detecting non-packed and packed programs in real-world scenarios?  
\end{itemize}

\diffccs{To answer RQ1, we design four detection scenarios to evaluate the performance of models and entropy metric
(\secref{sec:sec5evaluation_packeddata}),}
To answer RQ2, we employ model-unseen packed programs and low-entropy packed data as detection targets
(\secref{sec:sec5evaluation_packeddata} and \secref{sec:sec5evaluation_low_entropy}),
For RQ3, we conduct experiments to evaluate the impact of epochs and (enable/disable) components on detection performance
(\secref{sec:sec5ablation_study}),
For RQ4, we compare Pack-ALM with mature heuristic-based tools on the real-world samples and perform case studies to show the feasibility of detecting in-the-wild custom packers and adversarial samples
(\secref{sec:sec5evaluation_wild} and Appendix \ref{app:case_study1}, \ref{app:case_study2}).

\subsection{Experimental Settings}
\label{sec:sec5evaluation_datasets}
Our Pack-ALM model is built upon the RoBERTa-based architecture within the Fairseq framework~\cite{Ott2019}. The development environment utilized PyTorch 2.1.2, CUDA 12.1, and CUDNN 8.8.1. 
The model employs 12 self-attention layers with multi-head attention, and each layer incorporates 12 attention heads. The hidden layer utilizes the GeLU~\cite{Hendrycks2016} activation function, while the pooling layer leverages the tanh function. 
During pre-training and fine-tuning, the maximum input length is set to 512 tokens, and the batch size is set to eight.
We run all experiments on a testbed machine with Intel i7-13700K CPU (16 cores, 3.40 GHz), NVIDIA RTX 4090 24GB, 32GB RAM, and 4TB SSD storage.

\subsubsection{Datasets}
\diffccs{As most prevalent Windows/Linux malware and benign binaries target x86/x64 architectures, our evaluation focuses on this dominant ecosystem to validate our core methodology.} Our experiments use three strictly isolated datasets. No programs or packers overlap between these datasets. 
The specific compositions are detailed below:

\vspace*{2pt}
\noindent\textbf{1. Pre-training Dataset (\dset{PTD}) } The \dset{PTD} comprises the instructions disassembled from \done{4,207} non-packed benign programs. It consists of SourceForge~\cite{SlashdotMedia} downloaded programs, Linux platform programs, and programs randomly selected from the binary disassembly benchmark dataset~\cite{Pang2022}. These programs range in size from 2 KB to 190 MB and are preprocessed through linear disassembly, resulting in a total of \done{235,954,417} instructions. This dataset serves to pre-train the model on the real-world benign programs' organized instructions and data. 

\vspace*{2pt}
\noindent\textbf{2. Fine-Tuning Dataset (\dset{FTD}) }
We first constructed an instruction pool containing \done{1,058,973,634} instructions, linearly disassembled from \done{2,388} non-packed programs and \done{1,990} packed programs. 
The non-packed programs include Windows system applications and benchmark programs~\cite{Pang2022} not present in the \dset{PTD} dataset. 
The packed programs consist of samples generated directly by ten packers, as well as samples produced using low-entropy packing techniques~\cite{Mantovani2020}.
To establish ground truth for the instructions and packed data regions within these packed programs, we employed a combination of {dynamic instrumentation (Intel PIN~\cite{Luk2005}), static analysis (Disassembly), VirusTotal consensus, and manual verification.}
The \dset{FTD} dataset was randomly sampled from this instruction pool and comprises 40 million instructions.

\vspace*{2pt}    
\noindent\textbf{3. Test Dataset } Our test dataset comprises over 29,000 programs, encompassing a total of \done{10} million randomly sampled instructions.
\diffccs{The dataset integrates samples from reputable prior works~\cite{Mantovani2020,Li2023}, consists of the following components:}
(i) Packing dataset (\dset{T-PD})\footnote{\label{fn:footnote1} The packers and packed programs used in dataset \dset{T-PD} differ from those used in constructing dataset \dset{FTD}.} contains \done{1,090} packed program samples generated by ten sophisticated packers and samples generated by the low-entropy packing technique. It is used to evaluate the model's performance in detecting model-unseen packed data.
(ii) Non-packed dataset (\dset{T-NPD}), containing \done{803} non-packed samples, is employed to measure the model's false positive rate for model-unseen non-packed data.
(iii) Real-world sample dataset (\dset{T-WLD}) contains \done{9,828} non-packed benign samples from the source-forge dataset, 13,553 packed malware samples from the low-entropy dataset, and \done{3,776} standard-packed samples from the PackingData dataset~\cite{Packing-boxteam}. 

\diffccs{To effectively mitigate data leakage, our dataset employs rigorous program-level separation: no single packer or packed program is present in both training and testing sets. This also applies to different versions or derivatives of the same packer/program. This strict enforcement prevents our model from inadvertently learning program-specific traits rather than generalized patterns.}

Our evaluation includes packers representing a broad spectrum of complexity levels, ensuring comprehensive coverage of the diverse scenarios identified in prior research~\cite{Ugarte-Pedrero2015}. 
Specifically, our dataset encompasses all six types of packers (Type I to Type VI), categorized by increasing levels of complexity as defined in~\cite{Ugarte-Pedrero2015}.
To rigorously evaluate model generalizability, we enforced a strict separation between the training and testing datasets for each packer, thereby eliminating any potential data overlap. 
Further details are provided in Appendix~\ref{sec:packer-types}. 

Constructing binary disassembly benchmark datasets for normal programs presents significant challenges~\cite{Pang2022}. 
This task becomes even more complex when considering packed programs that employ diverse obfuscation techniques and irregular program structures. 
To address this, we adopt the approach outlined in recent research~\cite{Pang2022,Pei2021}, utilizing debugging information (e.g., PDB files) to extract instruction and data boundaries within the non-packed program. 
For packed programs, we combine static analysis with dynamic binary instrumentation to delineate the boundaries of instructions and packed data.

\subsubsection{Peer Tools for Comparison}
\label{sec:sec5peer-work}

{Note that Mantovani et al.~\cite{Mantovani2020} have systematically evaluated ML-based methods and demonstrated that existing ML-based approaches lack reliability to detect (adversarial) packed programs.}
We choose four open-source assembly language models, including XDA~\cite{Pei2021}, PalmTree~\cite{Li2021}, {jTrans~\cite{Wang2022}}, and Word2Vec~\cite{Mikolov2013}. 
XDA is a general-purpose instruction embedding model that takes bytes as input.  
It demonstrates strong generalization ability at distinguishing instructions from inline data and tolerates open-source obfuscators. 
PalmTree is another general-purpose model that takes disassembled instructions as input.
{jTrans is designed for binary code similarity detection, which employ jump-aware representation of binaries and a new pre-training task designed to capture control flow information effectively.}
Word2Vec is a traditional word vector model, that serves as a baseline that does not leverage the Transformer architecture. 
To establish a fair comparison, we retrained these four peer tools using their publicly available source code on our \dset{PTD}.
\diffccs{We employed identical hyperparameters to fine-tune all models on the same training set. We meticulously monitored loss curves to confirm that all models had converged, and selected each model's best-performing checkpoint on the validation set for testing.}
\diffccs{We also compare Pack-ALM with state-of-the-art methods, including the ML-based tool PackHero~\cite{packhero_2025}, signature-based tools (DIE~\cite{Horsicq} and PackGenome~\cite{Li2023}), as well as with entropy-based detection.}
The entropy threshold is set to the common value of 7.0, and the detection granularity is performed at both the program and window levels.

\subsubsection{Downstream Detection Tasks}
\label{sec:sec4downstream-detection-task}
Considering prior research did not explore the models' ability to recognize packed data and programs.
We design two downstream detection tasks for the first time:

\vspace*{2pt} 
\noindent\textbf{Packed Data Region Identification }
The models are tasked with classifying pseudo instructions and real instructions. 
{Note that packed programs contain substantial packed data and few unpacking routine instructions.
In accordance with the entropy detection window settings,
we set models' detection window size to \done{100} instructions to balance testing speed and accuracy.} This task evaluates the model's ability to distinguish between instructions and packed data, 
\diffccs{providing support for accurately identifying packed data and (obfuscated) unpacking routine instructions.}

\vspace*{2pt} 
\noindent\textbf{Packed Program (or Section) Identification } 
The models are tasked with distinguishing packed programs (or sections) from non-packed ones. This task evaluates the model's ability to identify samples containing packed data at a coarse-grained level, 
aiding researchers in rapidly identifying valuable packed samples for further in-depth analysis.

\diffccs{\subsubsection{Evaluation Metrics}
During the evaluation, we adopt established metrics used in classification tasks~\cite{packhero_2025,Li2023}: accuracy, precision (P), recall (R), F1-score, true/false positive rates (TPR/FPR), true/false negative rates (TNR/FNR), and detection coverage rate (DCR), as well as the running time (T). 
Please note that running time measurements reflect only the model inference time, excluding preprocessing and pre-training stages.}

\begin{table}
    \centering
    \caption{Evaluation results of each method under the task of instructions/data differentiation and native/packed data differentiation. ``Task A'' represents the task of classifying real and pseudo instructions, and ``Task B'' represents the task of classifying native and packed data from pseudo instructions. }
    \label{tab:tab4-total-task}
    \begin{adjustbox}{width=\linewidth}

    \begin{tabular}{lcccc|cccc}
    \toprule
             & \multicolumn{4}{c|}{Task A} & \multicolumn{4}{c}{Task B} \\ \cmidrule(lr){2-5} \cmidrule(lr){6-9} 
             & F1 [\%]  & P [\%]  & R [\%]  & T [s]  & F1 [\%]  & P [\%]  & R [\%]  & T [s]  \\ \midrule
    \rowcolor{gray!20}
    Pack-ALM &    \textbf{97.73} &  \textbf{98.44}         &   \textbf{97.03}   &      192 & \textbf{89.11} & \textbf{88.76} & \textbf{89.48}  &     124   \\
        {jTrans} &    97.57          &  98.17 &    96.97           &     249 & 79.85 & 83.58 & 76.44  &      165  \\

    \rowcolor{gray!20}
    XDA      &    94.49          &   96.64         &    92.43           &      195 & 79.00 & 87.21 & 72.20  &      129  \\
    PalmTree &    88.65          &  98.50 &    80.58           &     775 & 79.29 & 80.63 & 78.01  &      508  \\
\rowcolor{gray!20}
    Word2Vec &    90.61          &   97.09         &    84.94           &      10 & 79.43 & 78.73 & 80.14  &      4  \\
    \hline
    \rowcolor{gray!20}
    Entropy  &     - &    - &    - &       - & 44.39 & 91.62 & 29.30  &       3 \\ \bottomrule
    \end{tabular}
\end{adjustbox}
\vspace{-5mm}
\end{table}

\subsection{Packed Data Identification}
\label{sec:sec5evaluation_packeddata}
\diffccs{This section evaluates the ability of models to identify packed data regions,  a critical step in packed program detection. }
We first apply the models (i.e., Pack-ALM, jTrans, XDA, PalmTree, and Word2Vec) to classify real and pseudo instructions.
Then, we use models to detect the packed data region by classifying pseudo instructions.
The packed data in this experiment are extracted from the packed programs generated by model-unseen packers.

\subsubsection{Classification of Real and Pseudo Instructions}
This experiment evaluates the models' ability to distinguish instruction and (packed) data by classifying real and pseudo instructions. We randomly sampled a test dataset containing six million real instructions and six million pseudo instructions from the aforementioned \dset{T-PD} and \dset{T-NPD} dataset.
Real instructions are extracted from non-packed programs and the unpacking routines of packed programs.
Pseudo instructions are disassembled from native data and packed data, including low-entropy packed data.

From the results summarized in the ``Task A'' column of \tabref{tab:tab4-total-task}, we can find out that Pack-ALM demonstrably outperforms all other models.
We enable PalmTree to classify real and pseudo instructions by feeding the linear disassembled instructions.

\subsubsection{Classification of Packed and Native Data}

To evaluate the ability of each method to further differentiate between native and packed data from pseudo instructions, we employed a dataset, that contains 
four million instructions from native executables' data and four million instructions from packed data. 
The overall detection results are summarized in the ``Task B'' column of \tabref{tab:tab4-total-task}.

The experiment results demonstrate that Pack-ALM achieves significantly superior classification performance compared to other models. 
Pack-ALM outperforms XDA by \done{12.8\%, 1.78\%, and 23.93\%} in F1, precision, and recall scores, respectively.
The models use disassembled instruction as input (i.e., Pack-ALM, PalmTree, {jTrans}, and Word2Vec) to achieve higher recall than the XDA, which uses binary bytes as input.
{Furthermore, jTrans's jump-aware embeddings contribute to its superior performance in detecting unpacking routines compared to the other three benchmark models.}
The improved recall enhances the model's ability to detect most packed data samples.

\begin{table}
    \centering
    \caption{Detection accuracy results of each method under the task of instruction/data differentiation and native/packed data differentiation.}
    \label{tab:tab5-task1-acc}
    \begin{adjustbox}{width=\linewidth}

    \begin{tabular}{lcc|cc}
    \toprule
    & \multicolumn{2}{c|}{Real Instruction} & \multicolumn{2}{c}{Pseudo Instruction} \\ \cmidrule(lr){2-3} \cmidrule(lr){4-5} 
     & Norm. Prog.  & Unpk. Routines  & Nat. Data           & Pkd. Data           \\ \midrule
    \rowcolor{gray!20}
    Pack-ALM    &   \textbf{99.43\%}           &      \textbf{92.01\%}            &   \textbf{93.30\%} &        \textbf{91.26\% }  \\
    {jTrans} &    99.24\%          &  91.63\% &    90.75\%           &     86.12\% \\
    \rowcolor{gray!20}
    XDA         &   99.10\%           &      76.80\%            &   93.13\% &        86.36\%   \\
    PalmTree    &   94.03\%           &      63.69\%            &   88.24\% &        88.68\%   \\
    \rowcolor{gray!20}
    Word2vec    &   94.92\%           &      67.88\%            &   83.31\% &        89.27\%   \\

    \bottomrule
    \end{tabular}
\end{adjustbox}
\vspace{-4mm}
\end{table} 

\subsubsection{Deep Factor of Detection in Different Scenarios}
To gain a deeper understanding of the factors that influence each model's performance, we examine models' classification accuracy on different sample types.

\noindent\textbf{Classifying Real Instructions.} 
As revealed in the ``Real Instruction'' of \tabref{tab:tab5-task1-acc}, Pack-ALM performs on par with {jTrans,} XDA and PalmTree in recognizing real instructions from normal programs.
This suggests Pack-ALM's potential application in broader binary analysis tasks.
Notably, Pack-ALM demonstrates a significant advantage over other models in correctly identifying the unpacking routine instructions. 
This highlights the superior generalization ability of our model in recognizing obfuscated instructions. Conversely, jTrans, XDA, PalmTree, and Word2Vec tend to misclassify obfuscated instructions in unpacking routines as pseudo-instructions.

\vspace*{2pt} 
\noindent\textbf{Classifying Pseudo Instructions.} 
\tabref{tab:tab5-task1-acc}'s ``Pseudo Instruction'' shows that Pack-ALM outperforms other models on detecting native and packed data.
When identifying the packed data, the models use instructions as input perform better than XDA model, which use binary bytes as input.
It also revels that our preprocess method can help model to capture the features and distinguish native and packed data.

\colorpar{\noindent\textbf{Answer to RQ1:} \diffccs{Pack-ALM outperforms state-of-the-art models in real/pseudo instruction classification.} For example, Pack-ALM outperforms other models by up to 10.08\%, 23.93\%, and 12.8\% in terms of precision, recall, and F1 score, respectively.}


\subsection{Evaluating Model Performance with Low-entropy Packing Techniques}
\label{sec:sec5evaluation_low_entropy}

Bundt et al.~\cite{bundt2023blackbox} recently highlight that current state-of-the-art assembly language models remain vulnerable to a range of adversarial techniques, including binary obfuscation, encryption, injected dead code, 
and opaque predicates---all of which are commonly employed by the packers included in our evaluation.
Given that packed programs inherently utilize diverse obfuscation strategies, our evaluation provides comprehensive coverage of the adversarially modified binaries discussed in~\cite{bundt2023blackbox}. Unlike prior benchmarks, however, our adversarial setting emphasizes low-entropy attacks, which more accurately reflect the real-world challenges associated with packer detection.

To evaluate the effectiveness of each model in detecting adversarially packed samples, we conduct a detailed analysis of various low-entropy packing techniques. 
As described in prior research~\cite{Mantovani2020}, low-entropy encoding strategies include byte padding, encoding, mono-alphabetic substitution, and transposition (detailed in \appref{app:low_entropy}).
In this experiment, we exclude the byte padding technique from evaluation due to the easily detectable nature of long repetitive padding data.

\begin{table}
    \centering
    \caption{Detection accuracy results of each method on four low-entropy packing techniques.}
    \label{tab:tab6-low-entropy-acc}
    \begin{adjustbox}{width=\linewidth}

    \begin{tabular}{lc|c|c|c}
    \toprule

    & \multicolumn{4}{c}{Low-Entropy Packing   Techniques}       \\ \midrule
     & Transportation & Encoding & Mono-alphabetic & Poly-alphabetic \\ \midrule
    \rowcolor{gray!20}
    Pack-ALM    &  \textbf{92.45\%} &       \textbf{94.85\%}    &    \textbf{96.00\%} &    \textbf{96.40\%}  \\
    {jTrans} &    65.25\%          &  80.55\% &    82.10\%           &     88.25\% \\
    \rowcolor{gray!20}
    XDA         &  61.80\% &       89.10\%     &    90.10\% &    92.35\%  \\
    PalmTree    &  59.70\% &       91.35\%     &    91.30\% &    96.45\%  \\
    \rowcolor{gray!20}
    Word2vec    &  63.95\% &       94.45\%    &    93.70\% &    96.10\%  \\
    \rowcolor{gray!20}
    \hline
    Entropy     &  0.00\% &        0.00\%      &    0.00\%  &    0.00\%   \\\bottomrule
    \end{tabular}
\end{adjustbox}
\vspace{-3mm}
\end{table}

As shown in \tabref{tab:tab6-low-entropy-acc}, Pack-ALM demonstrates superior efficacy and robustness in recognizing low-entropy packed data.
Our findings indicate that {jTrans,} XDA and PalmTree demonstrate poor performance in identifying transportation-packed instructions. 
This limitation arises because the transportation technique rearranges the byte order of instructions, making the transportation-packed instructions 
appear similar to normal instructions but without correct semantics. The poor performance of {jTrans,} XDA and PalmTree highlights their vulnerability in 
recognizing pseudo instructions that closely resemble real instructions.

In detecting the other three types of low-entropy packed data---namely, encoding, mono-alphabetic substitution, and poly-alphabetic substitution---Pack-ALM and 
Word2Vec demonstrate superior performance compared to {jTrans,} XDA and PalmTree. Upon examining the packed data, we observe that Word2Vec benefits from its ability to 
capture co-occurrence patterns of repetitive tokens, which enhances its detection capabilities.
For example, the instructions disassembled from base32-encoded data contain repetitive sequences such as ``\code{push REG; dec REG; inc REG}.''

To further evaluate the models' performance, we examine their detection results on a real-world low-entropy sample that employs a combination of transposition and custom encoding techniques (detailed in \appref{app:case_study2}). 
As demonstrated in \figref{fig:09-case-study}, our experimental results demonstrate that Pack-ALM accurately classifies instructions and packed regions at \diffccs{a window level}, matching the performance of human experts. 
In contrast, XDA, {jTrans}, and PalmTree are susceptible to obfuscated instructions and low-entropy packed data.

\colorpar{\noindent\textbf{Answer to RQ2:} Pack-ALM outperforms other models in identifying the packed data from model-unseen packers and low-entropy packed programs. It reveals that Pack-ALM can be generalized to the sophisticated packer identification scenario.}


\subsection{Ablation Study}
\label{sec:sec5ablation_study}

To evaluate the effectiveness of Pack-ALM's components, we conducted a set of ablation study to achieve controlled-variable evaluation of model training gains. 
We use four configurations in this study: 
(i) \textbf{Pack-ALM} is with the siMLM model pre-trained from the preprocessed instructions.
(ii) \textbf{Pack-ALM-S} is the model using RoBerta's default MLM pre-training task based on the preprocessed instructions.
{(iii) \textbf{Pack-ALM-W} is the model without pretraining step.}
(iv) \textbf{Pack-ALM-NP} represents the pre-trained model without our instruction preprocessing step. 

\begin{figure}
    \centerline{\includegraphics[width=\linewidth]{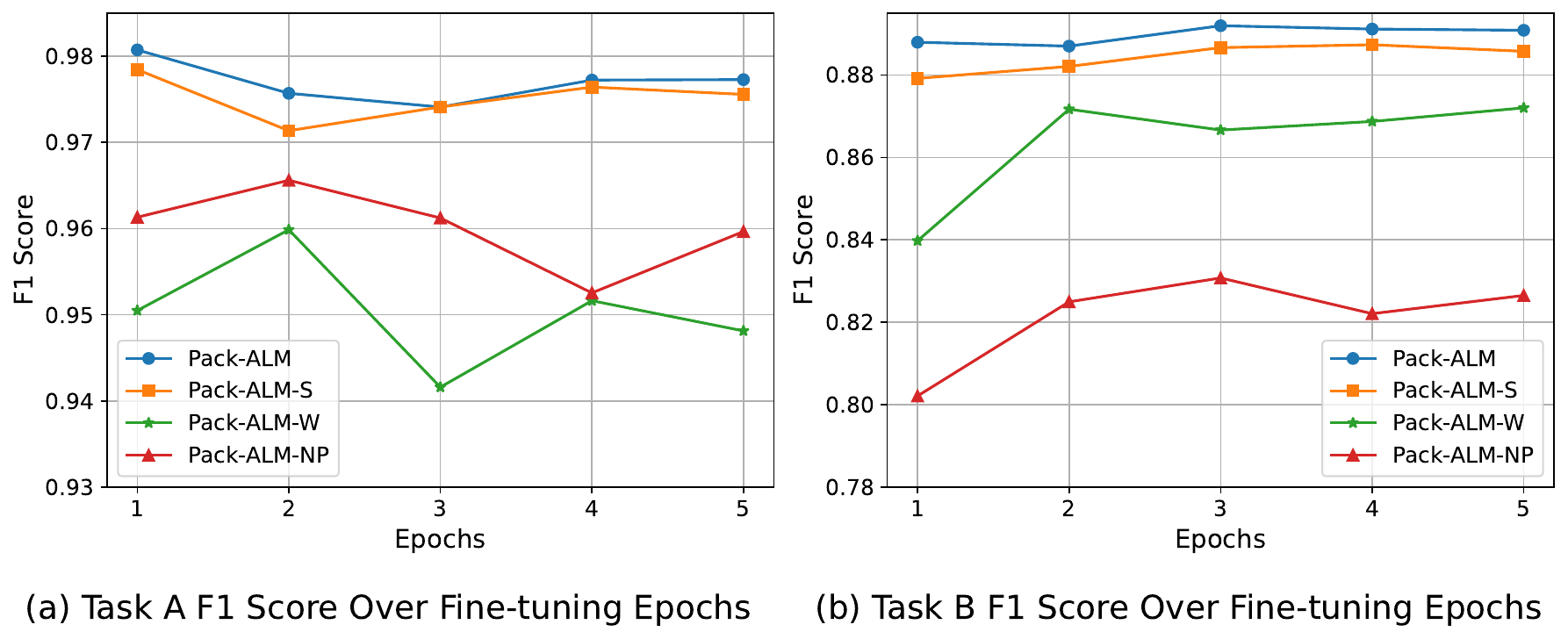}}
    \caption{\label{fig:05-ablation} Ablation experiment results of the Pack-ALM model. ``Pack-ALM-S'' means the model using RoBerta's default MLM pre-training task based on the preprocessed instructions. {``Pack-ALM-W'' means the model without pretraining step.} ``Pack-ALM-NP'' means the pre-trained model without our instruction preprocessing step.}
    \vspace{-3mm}
\end{figure}

\noindent\textbf{Instruction Preprocessing }
To evaluate the effect of our instruction preprocessing method on Pack-ALM's ability, we compared its with Pack-ALM-NP. 
Pack-ALM-NP treats each assembly instruction as a word and an entire segment as a sentence, aligning with prior research~\cite{Li2021,Chua2017}.
\figref{fig:05-ablation} demonstrates the significant performance improvements achieved by our instruction preprocessing method.
It improves Pack-ALM's F1 score by \done{1.34\%} in Task~A (i.e., real/pseudo instruction differentiation) and \done{6.85\%} in Task~B (i.e., native/packed data differentiation) on average.

\vspace*{2pt} 
\noindent\textbf{Pre-training Task siMLM }
To study the impact of our siMLM pre-training task, we compare Pack-ALM with Pack-ALM-S model.
By comparing the evaluation results in two subfigures of \figref{fig:05-ablation}, we observed that Pack-ALM achieve more benefit from siMLM pre-training task in Task~A than Task~B.
This improvement can be attributed to the siMLM pre-training task, which enables the model to learn the features of instruction's component.

To further investigate the impact of pretraining on downstream task performance, {we compare the Pack-ALM with Pack-ALM-W model.}
As shown in \figref{fig:05-ablation}, the pretrained model Pack-ALM demonstrated superior performance in both Task~A and Task~B compared to the non-pretrained baseline Pack-ALM-W. The performance gain was particularly pronounced in detecting adversarially packed data samples.

\vspace*{2pt} 
\noindent\textbf{Finetuning Epoch }
We further investigate the influence of the finetuning epoch on model performance.
Following the experimental setup in DeepDi~\cite{Yu2022}, we fine-tuned Pack-ALM for one to five epochs.
\figref{fig:05-ablation} presents evaluation results for four models on real/pseudo instruction classification tasks after different training epochs.
Pack-ALM achieves its peak F1 score during the first epoch, outperforming Pack-ALM-S and Pack-ALM-NP.
After being trained with more epochs, Pack-ALM achieves a similar F1 score at the fourth and fifth epochs.

\colorpar{\noindent\textbf{Answer to RQ3:} The ablation study shows that each component of Pack-ALM contributes to improving the performance. For example, our proposed preprocess method can improve the F1 score of Pack-ALM by \done{6.91\%} in native/packed data classification.}

\subsection{Evaluating Model Performance in the Wild}
\label{sec:sec5evaluation_wild}

\diffccs{Pack-ALM is designed to identify packed data.}
To extend Pack-ALM's capabilities to program-level detection, we train {a k-nearest neighbors (KNN) classifier and a convolutional neural networks (CNN) classifier} using Pack-ALM's output as input.
We preprocess Pack-ALM's output using N-gram analysis (n=3) and statistical feature extraction as input to train the KNN and CNN classifier.
The training dataset comprises \done{2,000} packed programs and \done{2,000} non-packed programs, randomly derived from the \dset{PTD} and \dset{FTD}. 
{Our trained Pack-ALM(K) (i.e., KNN classifier) and Pack-ALM(C) (i.e., CNN classifier) models are subsequently employed to categorize programs as packed or non-packed.}
This approach avoids relying on manually crafted heuristics and thresholds.

In our experiments, we choose DIE~\cite{Horsicq} and PackGenome~\cite{Li2023} as comparative signature-based detection tools, \diffccs{PackHero~\cite{packhero_2025} as the ML-based detection tool,} and use an entropy threshold of $\geq 7.0$ for entropy-based detection (detailed in \secref{sec:sec5peer-work}).
We conduct three experiments to evaluate the effectiveness of detection methods in the real-world scenarios, including non-packed programs, standard packer-packed programs, and adversarial packed programs.

\vspace*{2pt} 
\noindent\textbf{Non-packed Programs.}
This experiment evaluates the ability of each method to correctly identify non-packed programs.
The evaluation dataset comprises \done{9,828} non-packed programs randomly selected from the source-forge dataset of \done{\dset{T-WLD}}. We applied each method to the dataset, and the evaluation results are presented in \tabref{tab:tab8-wild-npd}. False positives indicate instances where the evaluation method mistakenly classified non-packed programs as packed.

Our evaluation results demonstrate that {Pack-ALM(K) and Pack-ALM(C)} achieves performance comparable to specific packer detection tools (i.e., DIE and PackGenome). 
{Pack-ALM(K)} only incorrectly classified \done{29} samples as packed, primarily due to the compressed (or encrypted) images and resources within the non-packed programs.
This low false positive rate is acceptable for generic packed program detection.
In contrast, entropy-based detection misclassified approximately $1,696$ non-packed programs as packed.
\diffccs{PackHero exhibits limitations in processing non-packed programs with complex and large call graphs (CGs), failing to analyze over 30\% of such cases due to model's constraints. This challenge further compromises its detection reliability for non-packed binaries.
}

\begin{table}
    \centering
    \caption{Detection performance of programs in the wild. {``Pack-ALM(K)'' and ``Pack-ALM(C)'' represent the program-level detection model, which comprises a KNN and CNN classifier trained on Pack-ALM's output, respectively}.
}
    \label{tab:tab8-wild-npd}
    \begin{threeparttable}
    \begin{adjustbox}{width=\linewidth}
    \begin{tabular}{lrrr|rrr|rrr}
    \toprule
                   & \multicolumn{3}{c|}{Non-packed} & \multicolumn{3}{c|}{Standard Packed} &
                   \multicolumn{3}{c}{Low-Entropy Packed} \\ 
                   \cmidrule(lr){2-4} \cmidrule(lr){5-7} \cmidrule(lr){8-10} 
                   & FPR [\%]  & TNR [\%] & DCR [\%] & FNR [\%]  & TPR [\%] & DCR [\%] & FNR [\%]  & TPR [\%] & DCR [\%]      \\ \midrule
    \rowcolor{gray!20}
    Pack-ALM(K)      &         0.30  &       99.70   & 100 &      0.95   &     99.05   & 100 &      4.68   &     95.32 & 100\\
    Pack-ALM(C)      &         5.28  &       94.72  & 100 &      0.53   &     99.47 & 100 &      9.16   &     90.84 & 100\\
    \rowcolor{gray!20}
    PackGenome     &         0  &       100 & 100  &    31.04    &   68.96& 100 &    96.49    &   3.51 & 100\\
    PackHero \tnote{1}     &         52.94  &       47.06  & 65.75 &    36.94    &   63.06 & 94.47 &    36.12    &   63.88 & 78.58\\
    \rowcolor{gray!20}
    Detect it Easy &         0  &       100 & 100  &      12.37   &     87.63 & 100 &      99.04   &     0.96 & 100\\
    \hline
    Entropy 7.0    &       17.26   &      82.74  & 100  &     23.65   &     76.35 & 100 &     99.97   &     0.03  & 100 \\
   \bottomrule
    \end{tabular}
    \end{adjustbox}
    \begin{tablenotes}
    \footnotesize
    \item[1] PackHero's dependence on program's call graph (CG) similarity for packer detection\\ impairs its effectiveness with large CGs, which preventing it from achieving 100\%\\ total detection coverage rate (DCR).
    \end{tablenotes}
    \end{threeparttable}
\end{table} 

\vspace*{2pt} 
\noindent\textbf{Standard Packed Programs.}
This experiment evaluates the ability of the Pack-ALM to identify standard packed programs. We apply {Pack-ALM(K), Pack-ALM(C)}, and other tools (DIE\diffccs{, PackHero} and PackGenome) to the \done{\dset{T-WLD}} dataset, comprising \done{3,776} packed programs generated by off-the-shelf packers. As shown in \tabref{tab:tab8-wild-npd}, {Pack-ALM(K) and Pack-ALM(C)} demonstrated superior performance, significantly outperforming DIE\diffccs{, PackHero} and PackGenome on the unseen packers.

Evaluation results show that both DIE and PackGenome exhibit diminished performance in detecting packed programs generated by \done{three} ``unseen'' off-the-shelf packers. 
After analyzing detection results, we observed that these tools struggle to identify packers not included in their signature databases. 
This limitation is inherent to all signature-based detection methods, as they rely on the maintenance and creation of signature rules for newly emerging packers.
\diffccs{Similarly, PackHero detects packed programs by identifying call graph similarities with known packed samples. However, its effectiveness significantly degrades when processing samples packed with novel, unseen packers.}
In contrast, Pack-ALM(K) achieves a true positive rate of \done{99.05\%} and a false negative rate of \done{0.95\%} in identifying packed programs.

\begin{table}
    \centering
    \caption{Overall detection performance of programs in the wild, using weighted average metrics across all categories to mitigate bias from individual results.
}
    \label{tab:tab8-wild-total}
    \begin{adjustbox}{width=\linewidth}
    \begin{tabular}{lrrrrr}
   \toprule
               & Precision [\%] & Accuracy [\%] & Recall [\%]  & F1 [\%]  & DCR [\%]    \\ \midrule
    \rowcolor{gray!20}
Pack-ALM(K)    & 99.83   & 97.43  & 96.13 & 97.95 & 100\\
Pack-ALM(C)    & 96.87   & 93.44  & 92.72 & 94.75 & 100\\
    \rowcolor{gray!20}
PackGenome     & 100  & 63.81  & 23.86 & 38.53 & 100\\
PackHero       & 67.96   & 57.68  & 63.70 & 65.76 & 76.15\\
    \rowcolor{gray!20}
Detect it Easy & 100  & 48.85  & 19.85 & 33.12 & 100\\\hline
Entropy 7.0    & 62.99   & 40.58  & 16.66 & 26.35 & 100\\ \bottomrule
    \end{tabular}
    \end{adjustbox}
\end{table} 

\vspace*{2pt} 
\noindent\textbf{Adversarial Packed Programs.}
Considering the malware authors often equip packed programs with adversarial techniques to evade detection.
To evaluate the method's ability to detect adversarial packed programs, we conduct experiments using adversarial samples from the \done{\dset{T-WLD}} dataset.
The evaluation results are summarized in \tabref{tab:tab8-wild-npd}. 
A true positive indicates the method correctly identified a packed sample. Conversely, a false negative occurs when an adversarial packed sample is mistakenly classified as non-packed.

Experimental results demonstrate that {Pack-ALM(K) and Pack-ALM(C)} consistently outperforms other tools. 
By comparing {Pack-ALM(K)} with signature-based detection tools, we find out that our model can identify an additional \done{12,552} samples, especially model-unseen packed programs.
Further analysis reveals that our model effectively mitigates the impact of obfuscated unpacking routines by focusing on packed data detection.
Notably, we identified \done{1,484} programs within the low-entropy dataset that were mistakenly labeled as non-packed. These programs actually contain packed data and exhibit unpacking behavior when executed in a sandbox environment (detailed in \appref{app:unexplore-packed-sample}).  

Moreover, compared to generic entropy-based packing detection methods, our model demonstrates significantly superior performance in detecting low-entropy packed programs.
We find that signature-based detection rules, which rely on detecting specific entry points and unpacking routines, can be easily evaded through targeted modifications to the packed binary.
\diffccs{Meanwhile, PackHero detects low-entropy packed programs by recognizing call graph similarities to traditional packed programs. However, it fails to recognize programs with complex call graphs.}

Another notable advantage of focusing on packed data detection is that our models can accurately identify incompletely unpacked samples. These samples are the unpacked (failed) samples that retain residual unpacking routine instructions but lack packed data (detailed in \appref{app:incomplete-unpacked-sample}).
These samples are first mentioned by PackGenome paper~\cite{Li2022}. Our experiment shows that these samples can greatly influence the signature-based packer detection tools like DIE and PackGenome, due to their reliance on scanning unpacking routine instructions.
In contrast, Pack-ALM's emphasis on packed data detection ensures its robustness against incompletely unpacked samples.

\vspace*{2pt} 
\noindent\textbf{Overall Detection Performance.}
\diffccs{To evaluate packer detection usability in real-world environments, we evaluated their detection performance on the whole dataset comprising non-packed, standard packed, and low-entropy packed programs.  
This methodology tests a tool's ability to directly distinguish program types and reflects real-world detection scenarios. 
As shown in \tabref{tab:tab8-wild-total}, Pack-ALM (K) and Pack-ALM (C) achieved the highest overall detection rates. 
Signature-based tools (i.e., PackGenome and DIE) exhibited high precision but low recall, indicating effectiveness limited to specific known packed programs. 
The ML-based PackHero demonstrated moderate performance, while entropy-based detection performed poorest due to significant degradation when encountering complex non-packed and low-entropy packed samples.}

\diffccs{
Furthermore, we examined each tool's ability to process samples normally. As shown in \tabref{tab:tab8-wild-npd}, Pack-ALM(K), Pack-ALM(C), DIE, and PackGenome achieved 100\% detection coverage rate, while PackHero exhibited significant limitations, achieving 65.75\%, 94.47\%, and 78.58\% for standard non-packed, standard packed, and low-entropy packed programs respectively.
This limitation stems from PackHero's call graph similarity matching approach, where large CGs impair detection effectiveness and prevent complete detection coverage.
}

\colorpar{\noindent\textbf{Answer to RQ4:} Pack-ALM can be extended to robustly classify real-world non-packed and packed programs. It outperforms entropy-based and signature-based industrial methods in identifying adversarial packed samples and incomplete-unpacked samples.}
\section{Discussion}
\label{sec:sec6discussion}

In this section, we reflect on the broader implications of our findings and examine the limitations of Pack-ALM in the context of real-world packed malware detection. We also explore how Pack-ALM can be extended to address challenges posed by complex obfuscation, cross-architecture compatibility, and fine-grained classification of packer families. Finally, we outline promising directions for advancing packing-aware assembly language models in future research.

\vspace*{2pt}
\noindent\textbf{Obfuscation Resilience of Assembly Language Models }
Existing research typically relies on lightweight obfuscation techniques (e.g., OLLVM~\cite{Junod2015}) to assess the obfuscation resilience of assembly language models~\cite{Pei2021,Jin2022}. 
However, real-world malware frequently employs more sophisticated techniques, such as binary packing, which significantly alter program structure and pose substantial challenges for static analysis. 
This study is the first to evaluate the effectiveness of assembly language models in handling commercial-grade obfuscation. Our experimental results show that Pack-ALM consistently outperforms existing models, 
owing to its novel instruction preprocessing strategy and tailored pre-training tasks. These findings suggest that Pack-ALM is well-suited for generalization to other binary analysis tasks involving complex obfuscation.

\vspace*{2pt}
\noindent\textbf{Complex Native Data and Packed Data }
Non-packed programs may contain complex data segments that resemble random data and could potentially confuse detection models. However, these segments are typically much shorter in length compared to packed data, allowing for effective differentiation. Our experimental analysis shows that packed data typically constitutes approximately \done{80\%} of a packed program's content, whereas complex native data in non-packed programs generally accounts for less than \done{10\%}.

\vspace*{2pt}
\noindent\textbf{Packer Classification }
Generic packer detection methods are effective in identifying previously unseen packed samples but often lack the granularity required to classify samples associated with specific packer families. To address this limitation, integrating Pack-ALM with established signature-based tools such as PackGenome~\cite{Li2023} and Detect It Easy (DIE) offers a promising solution. Researchers can first leverage Pack-ALM to detect packed programs and then extend its functionality to classify them based on similarities in the distribution patterns of real code, native data, and packed data. Subsequently, tools like PackGenome can be employed to identify specific packer families by analyzing the structural characteristics of their unpacking routines.

\vspace*{2pt}
\noindent\textbf{Impact of Different Architectures }
The proliferation of mobile and IoT devices has led to a rise in packed malware targeting multiple instruction set architectures (ISAs)~\cite{Duan2018, Kawakoya2023}, necessitating a comprehensive approach to cross-architecture packed malware detection. 
\diffccs{With appropriate training and pre-training datasets to Android/iOS/IoT binaries across diverse ISAs}, Pack-ALM has the potential to retain its detection effectiveness when migrated to new architectural targets. 
Additionally, cross-architecture models such as VulHawk~\cite{Luo2023}, which transform assembly instructions from diverse ISAs into customized intermediate representations (IR), 
offer a promising direction for architecture-agnostic analysis. As part of future work, we plan to extend Pack-ALM to support packed program detection on ARM and other widely used instruction set architectures.

\vspace*{2pt}
\noindent\textbf{LLMs for Binary Analysis }
\diffccs{Large Language Models (LLMs) are increasingly applied to binary analysis. 
Leveraging their advanced natural language understanding and generation capabilities, recent research~\cite{ullah_llms_2024,pearce_examining_2023,he_large_2023} has employed GPT-like models for tasks such as vulnerability comprehension and repair assistance.
However, LLMs lack inherent suitability for directly handling complex binary analysis tasks, such as automated vulnerability detection.
Due to their tendency to hallucinations and limited domain-specific knowledge in binary analysis, LLMs often require integration with Retrieval-Augmented Generation (RAG) systems and smaller specialized models to improve effectiveness in targeted domains. 
Pack-ALM's architecture targets efficient large-scale packed program detection through lightweight models.
This capability facilitates LLM-driven analysis in specific domains such as malware detection.} 
\section{Conclusion}

Entropy has been widely used as a metric for detecting packed programs. Traditional detection methods often depend on empirically determined high-entropy thresholds as a primary criterion. 
However, such approaches are easily circumvented by adversarial packing techniques (e.g., low-entropy packing). Similarly, signature-based detection methods, 
which rely on predefined empirical signatures, are susceptible to targeted adversarial packing techniques and face significant challenges in identifying novel packers.

To overcome the limitations of existing packer detection methods, we propose Pack-ALM, the first assembly language model specifically designed for packed program detection. 
Drawing inspiration from the real-word/pseudoword distinction in linguistics, we frame the challenge of packer detection as a task of distinguishing between real instructions and pseudo-instructions. 
Pack-ALM is trained to differentiate instructions from packed data by leveraging our preprocessed datasets of real and pseudo-instructions.
Extensive experimental results demonstrate that Pack-ALM surpasses entropy-based and signature-based detection methods, as well as peer assembly language models, in its resilience to adversarial techniques.
This advancement provides researchers with a more effective tool for accurately identifying packed programs and precisely \diffccs{locating packed data regions}.



\section*{Acknowledgments}

We sincerely thank ACM CCS 2025 anonymous reviewers for their insightful and helpful comments.
This work was supported by National Natural Science Foundation of China (62172238); National Key R\&D Program of China (2018YFA0704703). 
Jiang Ming was supported by NSF grants 2312185 \& 2417055 and Google Research Scholar Award.
Ni Zhang was supported by National Key R\&D Program of China for  ``Cyberspace Governance Special Fund'' (2021YFB3\\101701).

\bibliographystyle{ACM-Reference-Format}
\bibliography{sp25}

\begin{appendix}

\setcounter{table}{0}
\setcounter{figure}{0}
\renewcommand{\thetable}{A\arabic{table}}
\renewcommand{\thefigure}{A\arabic{figure}}
\section*{Appendix}

\section{Packers Employed for Evaluation} \label{sec:packer-types}
{
As demonstrated in~\tabref{tab:tab9-packers}, distinct packers were employed in the training and test sets to prevent data overlap, thereby mitigating potential biases that could compromise the model's evaluation. 
This configuration also enables an assessment of the model's ability to detect unknown packers not encountered during training, simulating real-world scenarios involving novel or unseen obfuscation techniques.
}

{The selected packers span diverse versions, release years, and complexity levels (denoted as ``Type'' in \tabref{tab:tab9-packers}), capturing the evolutionary progression of packing techniques and ensuring relevance to malware from varying time periods. 
Additionally, packers of varying complexity were incorporated to assess their impact on Pack-ALM's performance. Complexity here reflects both the intensity of instruction obfuscation and the sophistication of packed data processing workflows.
}

{
Our experimental results demonstrate that Pack-ALM achieves robust detection accuracy across heterogeneous packer types.
}


\section{On the Effectiveness of Entropy Metrics}
\label{sec:sec3entropy_eva}

As discussed in \secref{sec:sec2challenges_generic}, entropy-based generic packer detection suffers from two inherent limitations (\textbf{L1} and \textbf{L2}) 
due to the entropy metrics' inability to accurately capture the semantic feature of data content.
To systematically examine these limitations in real-world scenarios, we conducted a comprehensive evaluation on datasets comprising over 40,000 non-packed and adversarially packed programs: 
(i) non-packed programs dataset contains 6,038 system programs collected from Windows 7/11, as well as \done{22,108} open-source x86/x64 Windows programs of various sizes from Sourceforge~\cite{SlashdotMedia}. 
(ii) low-entropy packed programs dataset consists of \done{14,553} packed malware retrieved from the low-entropy dataset~\cite{Mantovani2020}.

During evaluation, we considered two primary entropy thresholds: $H(X) \geq 7.0$, commonly used in previous research, and $H(X) \geq 6.5$, employed by the state-of-the-art industrial packer detection tool DIE \cite{Horsicq}. 
Following classification methods from prior research \cite{Mantovani2020}, detection was conducted at three levels of scan granularity: whole program (``File''), section level (``Section''), and sliding window level (``Window''). 
We set the window size to 2,048, consistent with industry standards (e.g., Packing-box \cite{Packing-box} and DIE).

\begin{table}[hp]
    \centering
        \caption{{Packers employed for training and testing phases during the fine-tuning process.}}
        \vspace{-3mm}
            \label{tab:tab9-packers}
    \begin{adjustbox}{width=\linewidth}
    \begin{tabular}{lrrr}
        \toprule
         Packer & Version & Year & Type  \\ \midrule

         \rowcolor{gray!20}
\multicolumn{4}{l}{Training set} \\ \midrule
         FSG & 1.3 & 2002 & Type-III  \\ \hline
         MEW & 1.2 & 2004 & Type-I  \\ \hline
         ACProtect & 1.32,1.41 & 2004 & Type-IV  \\ \hline
         WinUpack & 0.31 & 2005 & Type-III  \\ \hline
         NSPack & 2.3,3.7,4.1 & 2005-2011 & Type-III  \\ \hline
         Kkrunchy & 0.23a & 2006 & Type-I  \\ \hline
         Expressor & 1.8 & 2010 & Type-III  \\ \hline
         Molebox & 4.3018 & 2010 & Type VI  \\ \hline
         MPRESS & 2.18,2.19 & 2010 & Type-I  \\ \hline
         ASPack & 2.29,2.38,2.42 & 2012-2017 & Type-III  \\ \midrule
                  \rowcolor{gray!20}
\multicolumn{4}{l}{Testing set} \\ \midrule
         UPX & 1.0,1.2,1.25,2.0,3.09,3.96 & 2000-2020 & Type-I  \\ \hline
         Armadillo & 6.04,7.00,8.00 & 2008-2010 & Type VI  \\ \hline
         Enigma & 1.55,3.1,3.8,4.2 & 2008-2014 & Type V  \\ \hline
         Zprotect & 1.6 & 2010 & Type VI  \\ \hline
         PE Compact & 3.022,3.11 & 2010-2012 & Type-III  \\ \hline
         VMProtect & 2.46,3.40 & 2010-2019 & Type VI  \\ \hline
         Obsidium & 1.5 & 2015 & Type-IV  \\ \hline
         Petite & 2.4 & 2015 & Type-III  \\ \hline
         Themida & 2.37,3.04 & 2015-2019 & Type-I\textasciitilde Type-III  \\ 
            \bottomrule
    \end{tabular}
    \end{adjustbox}
\vspace{-5mm}
\end{table}

As shown in \tabref{tab:entropy-detection}, our evaluation results demonstrate that entropy-based methods are ineffective in identifying non-packed and low-entropy packed programs.
For limitation \textbf{L1}, $H(X)\geq7.0$ threshold fails to detect low-entropy packed programs. 
Meanwhile, the $H(X)\geq6.5$ threshold incurs a high false positive rate on non-packed programs, with only limited improvement in detecting low-entropy packed samples. 
For example, the Windows program ``charmap.exe'' contains ``\code{.rsrc}'' and ``\code{.reloc}'' sections that are rich in complex data, leading to misclassification as packed when using the $H(X)\geq6.5$ threshold.
Regarding limitation \textbf{L2}, we find that empirically defined granularity can contribute to a higher false positive rate. 
For example, the non-packed program ``wordpad.exe'' stores a jump table within its ``\code{.text}'' section, which is misclassified as a high-entropy packed region at the window-level granularity. 

These findings underscore the need for more advanced approaches that can accurately identify adversarially packed malware while simultaneously reducing false positives.

\section{Case Study I: The Influence of Obfuscation on Unpacking Routine Recognition Processes}
\label{app:case_study1}

This case study analyzes the impact of obfuscation on the model's instruction recognition accuracy.
Packed data can hinder the recognition of unpacking routines within packed programs. 
Unpacking routines often employ sophisticated obfuscation to impede signature detection.
Meanwhile, they are frequently interweaved with packed data, further complicating the identification of real instructions.

\begin{table}
    \centering
    \caption{Entropy-based packed program detection method's effectiveness at different settings of  entropy threshold and detection granularity.}
    \vspace{-3mm}
    \label{tab:entropy-detection}

\begin{tabular}{lrr|rr}
\toprule
\textbf{Threshold}  & \multicolumn{2}{c}{$\mathbf{H}(X) \geq 7.0$ is Packed}  & \multicolumn{2}{c}{$\mathbf{H}(X) \geq 6.5$ is Packed} \\ \midrule
  & FPR [\%]     & FNR [\%]    & FPR [\%]        & FNR [\%]        \\
\midrule
\multicolumn{4}{l}{\textit{Windows  System Programs (non-packed)}} &\\
\rowcolor{gray!20}
File          & 2.00\%   & - & 19.89\%     & -    \\
Section        & 18.40\%  & - & 38.85\%     & -    \\
\rowcolor{gray!20}
Window         & 25.32\%  & - & 46.74\%     & -    \\
\midrule
\multicolumn{4}{l}{\textit{SourceForge Programs (non-packed)}}  &\\
\rowcolor{gray!20}
File          & 16.60\%  & - & 29.62\%     & -    \\
Section        & 28.27\%  & - & 45.40\%     & -    \\
\rowcolor{gray!20}
Window         & 42.51\%  & - & 59.44\%     & -    \\
\midrule
\multicolumn{4}{l}{\textit{Low-Entropy Packed Programs}} &\\
\rowcolor{gray!20}
File          & - & 99.97\% & -    & 71.44\%     \\
Section        & - & 99.97\% & -    & 44.29\%     \\
\rowcolor{gray!20}
Window         & - & 30.58\% & -    & 9.51\%     \\ \bottomrule
\end{tabular}
\vspace{-5mm}

\end{table}

\begin{figure}
\centerline{\includegraphics[width=\linewidth]{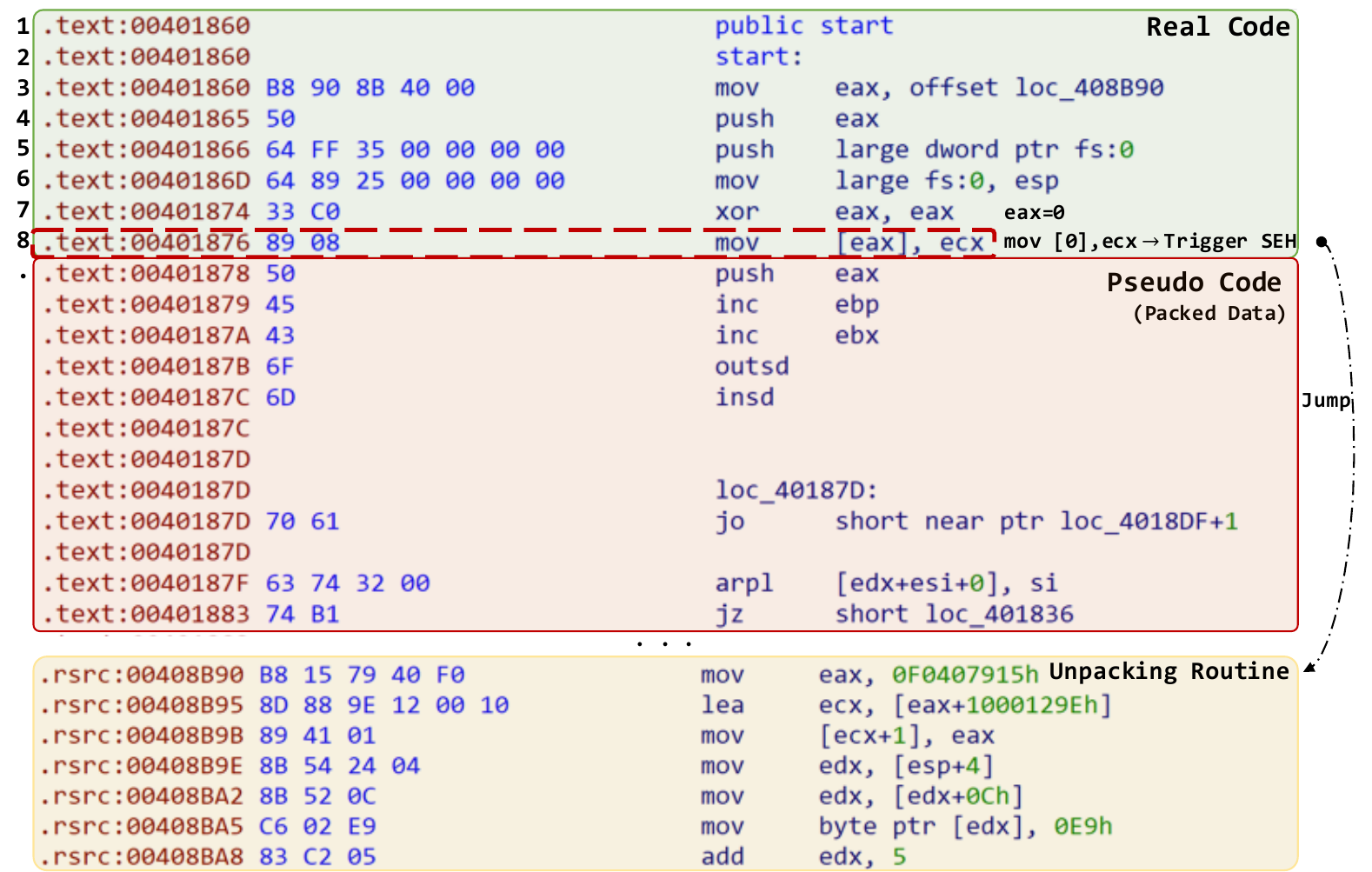}}
\caption{\label{fig:pecompact-ida} PECompact-packed program's entry point instructions disassembled by IDA.}
\vspace{-5mm}
\end{figure}

\figref{fig:pecompact-ida} is the PECompact-packed program's entry point instructions disassembled by IDA. 
The entry point instruction at the 8th line in \figref{fig:pecompact-ida} will trigger an exception.
It will lead to the structured exception handling (SEH) mechanism and ultimately jump to the real unpacking routine area within SEH.
It mistakenly interprets subsequent obfuscated data as instructions and produces approximately 109 incorrectly parsed instructions (red-colored pseudo code region in \figref{fig:pecompact-ida}), which will impede analyst and downstream analysis. 
The disassembly result demonstrates that even the commercial-grade disassembly tools like IDA can be hindered by obfuscation.
We also find out that both XDA and PalmTree models fail to recognize this region accurately.
In contrast, our Pack-ALM model accurately identifies such pseudo instructions as packed data.

\section{Superset Disassembling the Packed Programs}
\label{app:redundant-disassembly}
Superset disassembly disassembles the program's ``\code{.text}'' section at every byte offset.
Initially designed for binary rewriting, this approach has been widely adopted by many state-of-the-art disassembly methods (e.g., probabilistic disassembly~\cite{Miller2019} and DeepDi~\cite{Yu2022}) due to its advantages such as no false negatives.
However, superset disassembly has to process every section of the packed program, due to instructions and packed data are interweaves with the packed program (detailed in \secref{sec:sec2binary-packing}).
Since \done{80\%} of packed samples is the packed data, a substantial number of false positive instructions generated from these regions can significantly hinder binary analysis methods relying on superset disassembly.

To evaluate the redundant rate of superset disassembly on packed programs, we examine the ZProtected-packed sample as an example.
Superset disassembly exhibited a \done{99.5\%} false positive rate (FPR), with \done{97.5\%} of false positive instructions originating from packed data. 
This excessive redundancy significantly will significantly complicate subsequent analysis, such as control flow relation analysis used in probabilistic disassembly. 
Additionally, our experiment results reveal that IDA's false negative rate (FNR) is \done{91.9\%} and false positive rate (FPR) is \done{49.9\%}. 
It originates from IDA's limited disassembly strategy, which can be easily hindered by indirect jump (e.g., ``\code{jnz REG}'') obfuscations within unpacking routine.

\begin{figure}
    \centerline{\includegraphics[width=\linewidth]{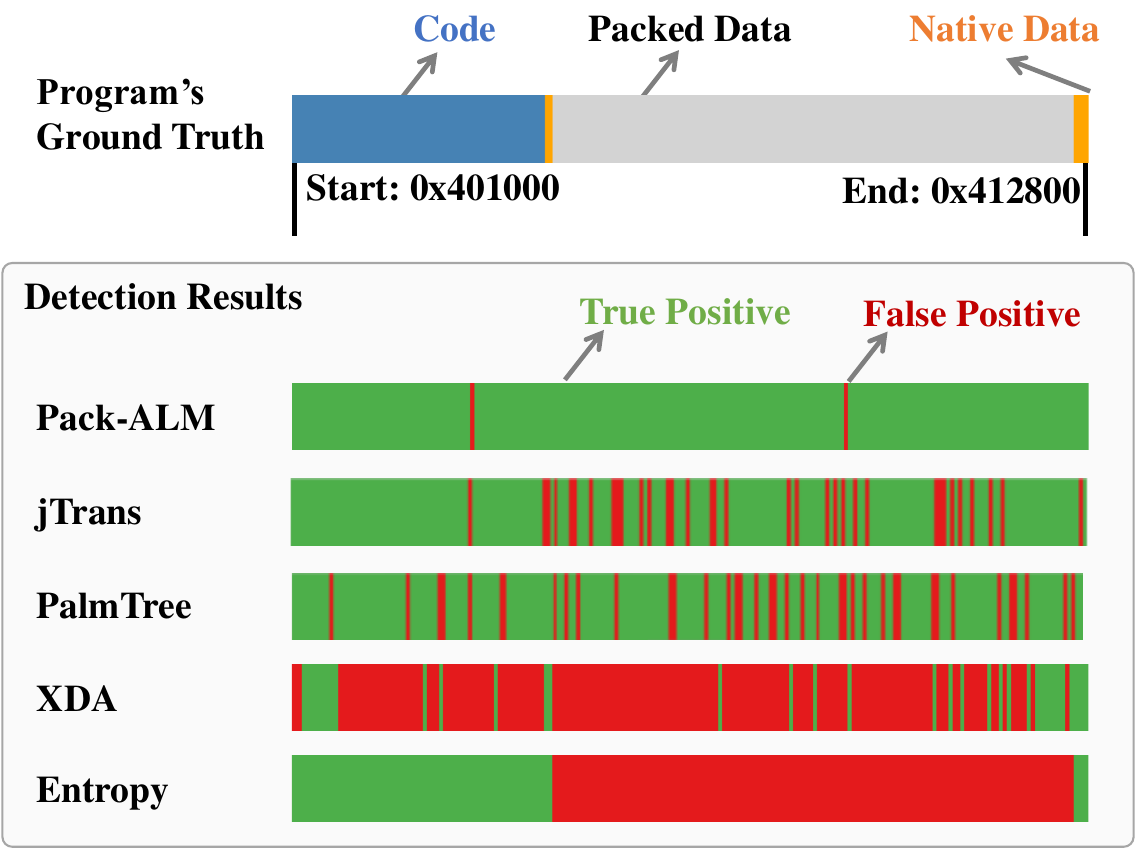}}
    \caption{\label{fig:09-case-study} An illustration of detection results for a low-entropy packed program. In the ground truth, blue regions represent code, gray regions represent packed data, and yellow regions represent native data. In the detection results, green regions indicate true positive detections, while red regions indicate false positive detections. }
    \vspace{-5mm}
\end{figure}
\section{Information Entropy and Low-Entropy Packing}
\label{app:low_entropy}

Entropy, a longstanding metric in binary packing detection, relies on empirically defined thresholds and detection ranges. 
For example, most studies adopt a threshold of H(X)$\geq$7.0 to identify packed programs.
However, the neglect of data element arrangement and reliance on empirical thresholds render these detection methods vulnerable to low-entropy binary packing techniques.
In this section, we explore the core reasons for entropy's ineffectiveness when encountering low-entropy packing and complex non-packed program scenarios.

\subsection{Calculation of Information Entropy}
In the packer detection scenario, the Shannon entropy $H(X)$ quantifies the uncertainty of a program's content. It is calculated using the formula:
\begin{equation}
H(X) = -\sum_{x\in \mathcal{X}}p(x) * \log_{2}p(x) 
\end{equation}
where $X$ is an input program or segment, $\mathcal{X}$ is a set of all possible byte values, $x$ is a specific byte value, and $p(x)$ is the probability of $x$ appearing in $X$.
Given that $\mathcal{X}$ ranges from 0 to 255, the entropy value falls between 0.0 and 8.0.


The entropy increases with data disorder. 
When each $p(x)$ equals 1/256, the uniform distribution of $p(x)$ results in the maximum entropy value of 8.0.
The lossless compression replaces repeated bytes with shorter representations, approximating a uniform distribution and consequently increasing entropy.
Meanwhile, encryption introduces uncertainty and randomness, also increasing entropy.
Since the packed program must contain substantial compressed (or encrypted) data, its high entropy can serve as an indicator of packed data. 

\subsection{Influence of Low-Entropy Binary Packing on Entropy}

Mantovani et al.~\cite{Mantovani2020} identified several strategies employed by packed malware to reduce program entropy and evade entropy-based detection:
\begin{itemize}
    \item \textbf{Byte Padding}. It adds substantial amounts of repetitive low-entropy data at specific locations within the program. For example, inserting a section filled with 0xff characters can decrease the entropy. This increases the $p(x)$ of certain bytes, therefore reducing overall entropy. 
    \item \textbf{Encoding}. It uses a different symbolic alphabet (e.g., base64, base32, or a custom encoding algorithm) to encode the data, which can reduce entropy by decreasing the size of the $\mathcal{X}$. For example, base64-encoded data's $\mathcal{X}$ is $[0,64]$, limiting the maximum entropy is 6.0.
    \item \textbf{Mono-alphabetic Substitution}. Using a substitution algorithm to shift each byte of the data, does not alter the probability distribution of $p(x)$ and therefore does not change the entropy.
    \item \textbf{Transposition}. It rearranges the byte order within the program by switching the byte value. Since the $p(x)$ of each byte is unchanged, the entropy of the rearranged data remains unchanged. 
    \item \textbf{Poly-alphabetic Substitution}. Similar to the Virginia cipher, it employs multiple alphabets for sequential substitution, and can slightly increase entropy due to changes in the $p(x)$ distribution.
\end{itemize}
By manipulating either the $\mathcal{X}$ or the probability distribution $p(x)$, these low-entropy packing techniques can effectively minimize the entropy of packed program, posing a significant challenge to entropy-based detection.

\begin{figure}[htp]
\centering
    \begin{tabular}{c}
    \begin{lstlisting}[style=customc, linewidth=0.8\linewidth, autogobble=true]
 // Transposition
 PACKED_START = 0x408048
 PACKED_RANGE = 0x6926
 PACKED_END = PACKED_START + PACKED_RANGE 
 DICT_START = 0x40e970 // DICT_RANGE = 0x100
 for addr in [PACKED_START, PACKED_END]{
   offset = ReadByte(addr)
   decoded_byte = ReadByte(DICT_START + offset)
   DEST <@$\leftarrow$@> WriteMemory(decoded_byte)
 }
 // Custom Encoding
 for index < PACKED_RANGE {
   offset_byte = DEST[index++]
   if (offset_byte == DEST[0]){
     if (DEST[index]){
       UNPACKED_DEST <@$\leftarrow$@> Decode(DEST[index], index)
       }else{
         UNPACKED_DEST <@$\leftarrow$@> DEST[0]
         }	
   }else{
     UNPACKED_DEST <@$\leftarrow$@> offset_byte
 }
    \end{lstlisting}
    \end{tabular}
    \caption{\label{fig:simplified-unpacking} The simplified unpacking algorithm of a low-entropy packed sample.}
 \vspace{-3mm}
\end{figure}
\section{Case Study II: Identifying Adversarial Packed Data}
\label{app:case_study2}
Identifying adversarial packed data remains a significant challenge due to malware developers' adoption of adversarial strategies to conceal program characteristics and evade antivirus engines. 
The robustness of detection methods for adversarial packed data identification is paramount. Low-entropy techniques, as highlighted by Mantovani et al.~\cite{Mantovani2020}, represent the most effective means of bypassing general entropy-based detection techniques. 
These techniques typically avoid introducing random data with distinctive features like compression or encryption algorithms, instead producing data with a specific regular distribution. 
This poses a substantial challenge for entropy detection, which identifies packed data based on data distribution.

To illustrate this challenge, we examine a low-entropy packing sample 
(SHA256: 34b2bbc691da589a36a2f09e67219b4041d29dfff86\\8a4430bc75b0e3e62276e) 
randomly selected from the low-entropy dataset~\cite{Mantovani2020}. 
Through reverse engineering, we identified that this sample contains low-entropy packed data using a combination of transposition and custom encoding techniques to reduce entropy. 
This program stores \code{0x6926} packed data in the ``\code{.data}'' section and is unpacked using a simplified algorithm in \figref{fig:simplified-unpacking}.
We employ each model to identify the custom-packed data within this low-entropy sample.
\figref{fig:09-case-study} illustrates the performance of each model in classifying the packed data region.
Red regions indicate incorrect classifications, while green regions denote correct classifications.
Pack-ALM exhibits superior detection effectiveness in identifying transposition-packed data.
PalmTree with an accuracy rate of only \done{80.98\%}, likely because it can only learn pairwise instruction relationships.
XDA's performs even worse, with an accuracy rate only is \done{18.54\%}, potentially due to its reliance on memorizing byte distribution patterns without verifying the syntactic correctness of switched byte sequences.

\begin{figure}
\centerline{\includegraphics[width=\linewidth]{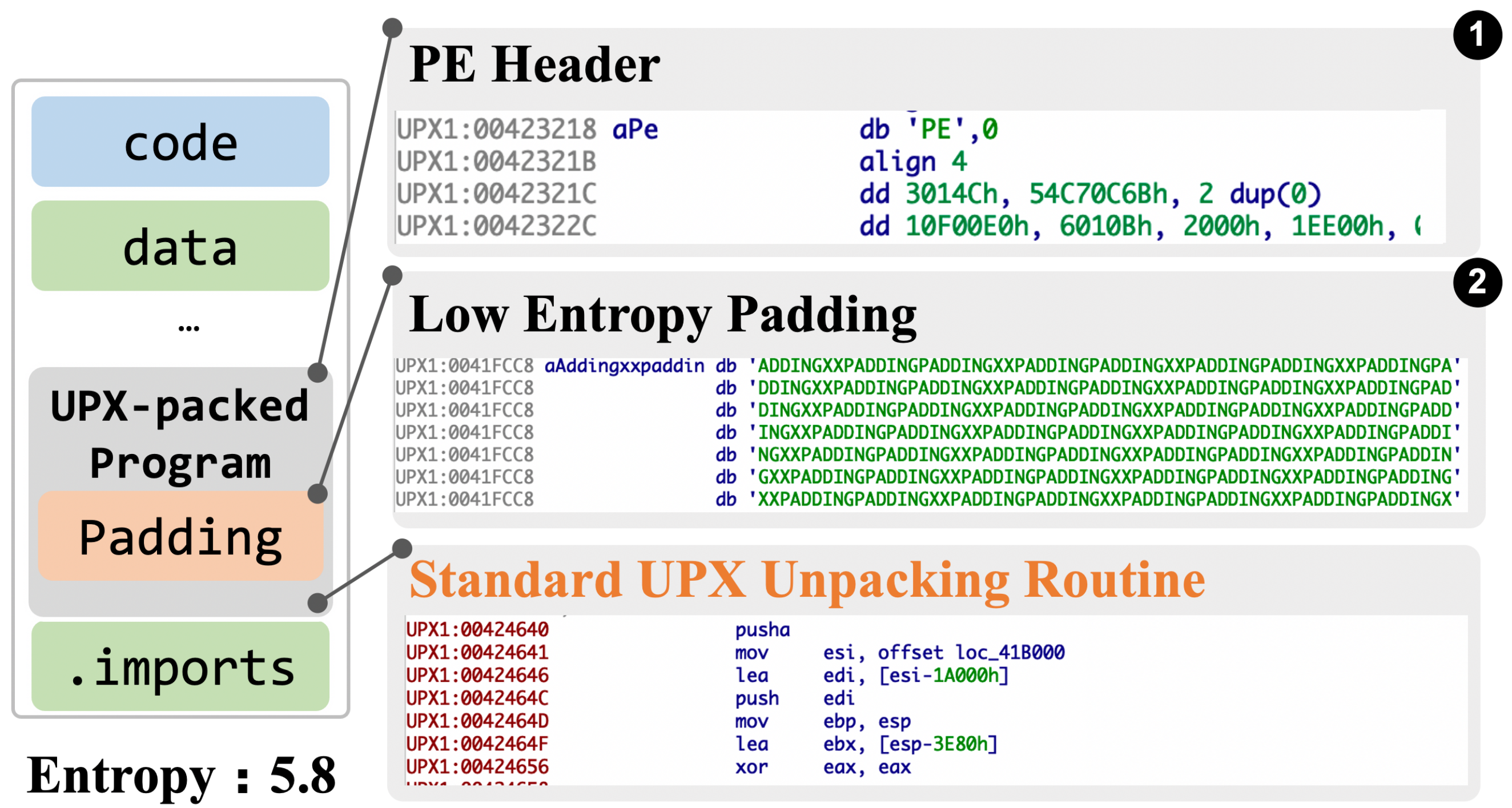}}
\caption{\label{fig:low-entropy-upx} An example of misclassified non-packed sample which conceal a UPX-packed program.}
\vspace{-4mm}
\end{figure}
\section{Unexplored Packed Samples in Low-entropy Non-packed Dataset}
\label{app:unexplore-packed-sample}
Prior dynamic-based detection methods~\cite{Ugarte-Pedrero2015,Mantovani2020} often focuses on detecting the classical ``written-then-executed'' feature, which release and execute unpacked code within the same process.
However, to evade detection, malware often employs sophisticated obfuscation techniques to camouflage unpacking routines.
Consequently, prior methods struggle to detect the evasive unpacking routine that decrypt malicious code into a new process or the hard driver~\cite{Or-meir2019}.
To investigate the impact of these evasive techniques, we apply Pack-ALM to scan \done{26,326} non-packed programs of Mantovani et al's low-entropy dataset \cite{Mantovani2020}.

Our experiments reveal that Pack-ALM successfully identifies packed data in approximately 24.3\% of programs labeled as native data.
After manually inspecting the samples, we find out that many of these packed samples were likely overlooked by Mantovani et al.'s dynamic packer detection tool.
For example, the sample (SHA256: \done{a0ab007114054de8932a6e2b7b49d9a726353506fac8fbb19f\\4633cf06ca1621}) conceals a UPX-packed program starting at address \code{0x423218} (\circled{1} in \figref{fig:low-entropy-upx}).
To reduce entropy, this sample employs low-entropy padding from \code{0x41F4C8} to \code{0x41FCC8} (\circled{2} in \figref{fig:low-entropy-upx}).
Existing detection tools, such as the industrial tool DIE, misclassify this sample as non-packed based on either signature rules or entropy strategy. 
In contrast, PackGenome accurately detects the sample by recognizing unpacking routine instructions. 
Meanwhile, the online sandbox service can also capture and identify the released UPX-packed program (``jhtfh.exe'').

\section{Incomplete-Unpacked samples}
\label{app:incomplete-unpacked-sample}
Incomplete-unpacked sample is the unpacked (failed) sample that typically consist of the unpacked original program and residual unpacking routines. 
These samples often arise from incomplete unpacking functionality of unpackers or dumped snapshots of the program's memory. 
While the residual unpacking routine's functionality may be disabled, these instructions can still be captured by static detection methods like signature-based detection.
Because static detection methods are limited to statically scanning packing-related features and cannot verify whether the residual unpacking routine is executed. 
To effectively identify incomplete-unpacked samples, we use the combined approach of signature-based detection and Pack-ALM.

\begin{figure}
\centerline{\includegraphics[width=\linewidth]{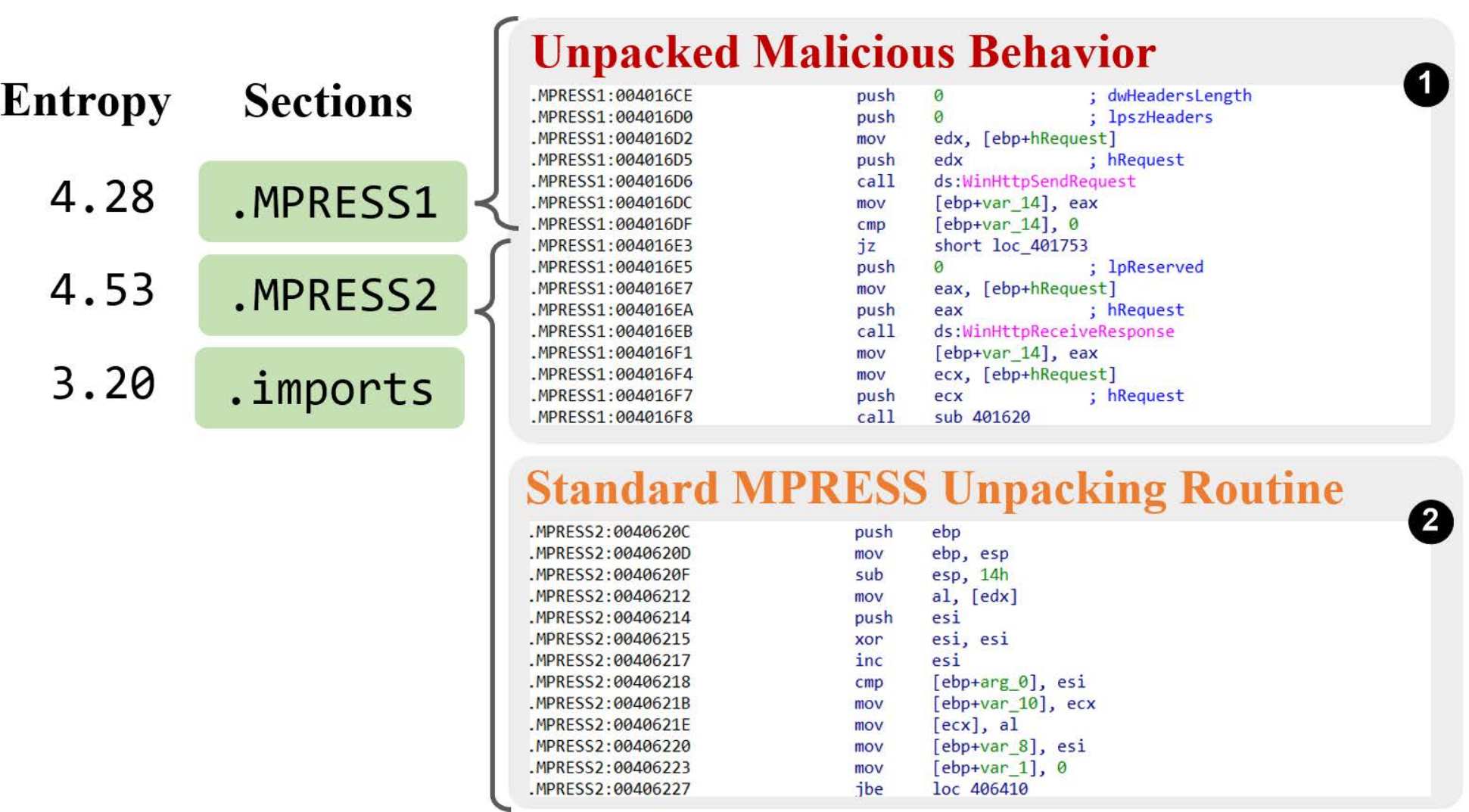}}
\caption{\label{fig:incomplete-unpacked} An example of incomplete-unpacked sample with UPX unpacking routine.}
\vspace{-4mm}
\end{figure}

We use the sample (SHA256:
21a6632db5d5749e598260927523591\\7783e6dd17f55fcc8c99b9930ddd07e0f) as the example.
This sample was falsely identified as MPRESS-packed by DIE and PackGenome.
\figref{fig:incomplete-unpacked} illustrates the program structure and corresponding entropy values.
In this sample, the compressed data have already been unpacked into normal code and data (\circled{1} in \figref{fig:incomplete-unpacked}), while the original unpacking routine (\circled{2} in \figref{fig:incomplete-unpacked}) remains. 
The residual unpacking routine and the ``\code{.MPRESS}'' section name misled  DIE and
PackGenome into classifying this sample as packed.

\end{appendix}


\end{document}